\documentclass[preprint,5p,twocolumn,nonatbib]{elsarticle_nonatbib}

\makeatletter
\let\c@author\relax
\makeatother

\expandafter\let\csname ver@natbib.sty\endcsname\relax

\usepackage[english]{babel}
\usepackage[utf8]{inputenc}
\usepackage{amsmath}
\DeclareMathOperator{\atantwo}{atan2}

\usepackage{graphicx}
\usepackage[colorinlistoftodos]{todonotes}
\usepackage{float}
\usepackage{hyperref}
\usepackage{filecontents,hyperref}
\usepackage[style=authoryear,backend=bibtex,doi=false,isbn=false,url=false,natbib=true]{biblatex}
\usepackage{csquotes}
\usepackage{mdframed}
\usepackage{textgreek}
\usepackage{array}
\usepackage{tabularx}
\usepackage{tabu}
\usepackage{boldline}
\usepackage{enumitem}
\usepackage{verbatim}
\usepackage{booktabs}
\usepackage{caption}
\usepackage{subcaption}
\usepackage{multicol}% http://ctan.org/pkg/multicols
\usepackage{url}
\usepackage{soul}

\usepackage{tikz}
\usepackage{tikzscale}
\usepackage{rotating}

\usepackage{circuitikz}
\tikzset{
	declare function={% in case of CVS which switches the arguments of atan2
		atan3(\a,\b)=ifthenelse(atan2(0,1)==90, atan2(\a,\b), atan2(\b,\a));},
	kinky cross radius/.initial=+.125cm,
	@kinky cross/.initial=+, kinky crosses/.is choice,
	kinky crosses/left/.style={@kinky cross=-},kinky crosses/right/.style={@kinky cross=+},
	kinky cross/.style args={(#1)--(#2)}{
		to path={
			let \p{@kc@}=($(\tikztotarget)-(\tikztostart)$),
			\n{@kc@}={atan3(\p{@kc@})+180} in
			-- ($(intersection of \tikztostart--{\tikztotarget} and #1--#2)!%
			\pgfkeysvalueof{/tikz/kinky cross radius}!(\tikztostart)$)
			arc [ radius     =\pgfkeysvalueof{/tikz/kinky cross radius},
			start angle=\n{@kc@},
			delta angle=\pgfkeysvalueof{/tikz/@kinky cross}180 ]
			-- (\tikztotarget)}}}

\tikzset{test/.style={font=\sffamily\fontsize{6pt}{6pt}\selectfont}}

\bibliography{mybibfile}

\usepackage{mathtools}

\let\oldnorm\norm   % <-- Store original \norm as \oldnorm
\let\norm\undefined % <-- "Undefine" \norm

\DeclarePairedDelimiter\abs{\lvert}{\rvert}%
\DeclarePairedDelimiter\norm{\lVert}{\rVert}%

\makeatletter
\let\oldabs\abs
\def\abs{\@ifstar{\oldabs}{\oldabs*}}
\let\oldnorm\norm
\def\norm{\@ifstar{\oldnorm}{\oldnorm*}}
\makeatother

\usepackage{graphicx}
\usepackage{xcolor}

\renewcommand{\Re}{\mathbb{R}}

\usepackage[]{algorithm2e}

\makeatletter
\newcommand{\removelatexerror}{\let\@latex@error\@gobble}
\makeatother

\usepackage{amssymb}
\usepackage{nomencl}
\makenomenclature
\usepackage{etoolbox}
\renewcommand\nomgroup[1]{%
	\item[\bfseries
	\ifstrequal{#1}{A}{Physical Constants}{
		\ifstrequal{#1}{B}{Variables}{
			\ifstrequal{#1}{C}{Ocean model}{
				\ifstrequal{#1}{D}{Oil model}{
					\ifstrequal{#1}{E}{Estimated Coefficients}{
	}}}}}
	]}
\begin{document}
	\begin{frontmatter}
		\journal{Marine Pollution Bulletin}
		\title{A combined ocean and oil model for model-based adaptive monitoring\tnoteref{t1}}
		\author[1]{Zak Hodgson\corref{cor1}}
		\ead{zhodgson1@sheffield.ac.uk}
		
		\author[2]{David Browne}
		\author[1]{I$\tilde{\text{n}}$aki Esnaola}
		\author[1]{Bryn Jones}

		\cortext[cor1]{Corresponding author. Tel.: +447510 168575.}
		\tnotetext[t1]{This document is a result of a research project funded by the University of Sheffield, the Engineering and Physical Sciences Research Council (EPSRC) UK and Andrew Moore \& Associates Ltd.}
		
		\address[1]{Department of Automatic Control and Systems Engineering, University of Sheffield, Sheffield, South Yorkshire, S1 3JD, United Kingdom}
		\address[2]{Andrew Moore \& Associates, 2703 Universal Trade Centre, 3 Arbuthnot Road, Central,	Hong Kong (SAR), China}
			
		\begin{abstract}
		This paper presents a combined ocean and oil model for adaptive placement of sensors in the immediate aftermath of oil-spills. A key feature of this model is the ability to correct its predictions of spill location using continual measurement feedback from a low number of deployed sensors. This allows for a model of relatively low complexity compared to existing models, which in turn enables fast predictions. The focus of this paper is upon the modelling aspects and in-particular the trade-off between complexity and numerical efficiency. The presented model contains relevant ocean, wind and wave dynamics for short-term spill predictions. The model is used to simulate the 2019 Grande America spill, with results compared to satellite imagery.  The predictions show good agreement, even after several days from the initial incident. As a precursor to future work, results are also presented that demonstrate how sensor feedback mitigates the effects of model inaccuracy.
		\end{abstract}
		
		\begin{keyword}
			Adaptive monitoring \sep Oil modelling \sep Contaminant monitoring
		\end{keyword}
		
	\end{frontmatter}
	%\maketitle
	
	\section{Introduction}
	The clean-up operations and legal claims that surround the annual average of 3500 \parencite{EMSA2018} maritime incidents and their lost cargo suffer from a lack of information, particularly in a remote location or in the immediate aftermath when current surveillance resources are unsuitable, unreliable or unavailable. While following a downward trend, around 6000 tonnes of oil are currently lost per year, with a 24 year peak of 116,000 tonnes in 2018 \parencite{ITOPF2019}. Clean-up operations, accident monitoring and rescue attempts are often hindered by the resources available at the accident locale, with specialist equipment including observation aircraft not arriving until several days after the event. Current observation solutions include satellites with Synthetic Aperture Radar (SAR) or other spectrum sensors and sensor-equipped vehicles although only aircraft or helicopters have the range and speed necessary to observe a large marine area quickly.
	
	Satellite data availability is limited to first responders and the most frequent sensor, SAR, is incapable of measuring oil thickness \parencite{Fingas2014} and the complex interplay between oil thickness, viscosity and wave parameters \parencite{Zhang2015a} result in further uncertainty in measurement results. SAR is unreliable in calm or rough seas (wind speeds less than 3m/s or greater than 10m/s) and environmental phenomena can produce false positives \parencite{Topouzelis2016}. Plausible incident sites must be verified by direct observation, usually meaning aerial observation.
	
	However, due to remoteness, flyovers are often conducted using local aircraft with no specialist sensors or tools and crewed by a human observer \parencite{ITOPF2014}. In extreme locations aerial observations are hampered by a lack of runways, requiring the chartering of vessels equipped with a helicopter pad, delaying observation by days, if-not weeks \parencite{Laruelle2011}. The expense of aircraft limits their number and hence the availability of simultaneous viewpoints or constant coverage during pilot/refuel breaks. Furthermore, health and safety concerns for the crew can limit their night-time deployment and their flight route is often pre-determined before take-off, with changes at the discretion of safety and airspace concerns. Observation aircraft plan routes as ladder search patterns in the supposed direction of wreckage or oil migration, with their data relayed back to model operators that correct oil locations manually. Despite their problems, aircraft offer excellent spatial and temporal resolution in their data and are exceptionally useful when directed by an oil drift model.
	
	Supporting tools, such as oil models, may not be available (due to a lack of data or resource allocation) in the crucial first few days following an incident. Existing models results provide useful data for response planning, but can require considerable time to do so, owing to their high complexity. However, despite their supposed accuracy, such model predictions still have to be verified by observation before resource allocation can commence\parencite{ITOPF2014}.
	
	The advent of increasingly low-cost autonomous platforms such as unmanned aerial vehicles (UAVs) and unmanned surface vehicles (USVs) offers the potential to sample an oil spill in a more rapid fashion and in greater number than the conventional methods listed above. However, what is currently lacking is the autonomy to direct such sensor platforms to the best spatial locations for sampling the spill, for the purpose of estimating its spatial extent. It is the opinion of the authors that such autonomy should be based upon a model that incorporates the key physical processes by which oil spreads upon the ocean surface, and that such a model should be `lean' in order to facilitate fast predictions for the purposes of real-time data assimilation from,  and subsequent issuing of guidance commands to, the mobile sensors. This paper describes the development of such a model, henceforth named the Sheffield Combined Environment Model (SCEM).
		
	A recent review of oil spill modelling \parencite{Spaulding2017} covers OSCAR \parencite{Reed2000}, SIMAP/OILMAP \parencite{FrenchMcCay2016}, GNOME/ADIOS \parencite{Lehr2002}, though other notables in the field include the model MEDSLIK \parencite{DeDominicis2013a} and work in support of hydrodynamic models.
	
	The review affirms modern oil spill models are complex amalgamations of Lagrangian (particle based) transport processes and varied algorithm types (stochastic and deterministic) of other processes, such as entrainment in the water column, or evaporation. There are some exceptions that use an Eulerian approach \parencite{Taylor2003}, but these are more limited in scope since supporting algorithms (such as entrainment) are Lagrangian based \parencite{Wang2010}, providing solutions per particle. State of the art 3D models aim to provide the most accurate estimations possible of oil position/properties, both surface and sub-surface, at the expense of computational speed, over an extended period of time (weeks to months of prediction) and hence include weathering effects. Inputs to the models, including geographical, wind and water current data, all come from exterior hydrodynamic models, that also vary in approach.
	
	Environmental models provide the oil spill model with wind, wave and current velocity data. Modern 3D models commonly use a harmonic water-level tide model for boundary in-flows and out-flows and base their physical processes and turbulence closure on the work of Mellor \parencite{Mellor1982}, \parencite{Mellor2003}. This includes 3D Navier Stokes, radiation stress from linear surface waves and a Smagorinsky eddy parametization, but with differing discrete solution methods such as an unstructured mesh \parencite{Wang2010}. Continuing work enabled coupling the wave model with an ocean model, and modification to incorporate depth-induced wave-breaking and wave-current interaction \parencite{Mellor2008}. Wave models are still external to the ocean model in most cases \parencite{Spaulding2017}, with the notable exception of Mellor's continuing research, a joining of the Stevens Institute of Technology Estuarine and Coastal Ocean Model (sECOM) and Mellor-Donelan-Oey (MDO) wave model \parencite{Marsooli2017}. Some work omits Ekman currents completely (slow forming horizontal net water currents due to the force balance between the coriolis effect and wind shear), others prefer to account for them (instantaneously forming) in their oil drift angle formulation \parencite{DeDominicis2013a}, while others include them in their 3D hydrodynamic model by including a coriolis force term in the Navier-Stokes equations \parencite{Marsooli2017}. For SCEM, consideration was given to utilisation of the Regional Ocean Modelling System (ROMS) \parencite{Moore2011} but it was decided overly complex given the focus of SCEM on model-guided sensing rather than long term predictions. 
	
	Due to constraints on computation, communication and time, the current 3D hydrodynamic models are unsuitable for sensor guidance applications. For example, a state of the art 3D model can take 74 hours to solve a 9 day simulation across 66000 nodes (the most useful measure of area), or approximately 400km x 300km, on an 8 CPU OpenMP computer \parencite{Marsooli2017a}. By contrast, for the purposes of rapidly mapping a spill in the~\emph{immediate aftermath} of a spill, a model need only make accurate predictions a few hours ahead, hence reducing the need for high complexity. This required complexity can be further lowered if the model predictions are continually corrected using repeated re-calculation based upon the most recent sensor information; a well-established technique known as `state-estimation' within the control systems community. Table \ref{tab:oilmodeltable} displays a summary of oil models, their internal environment models, intended prediction horizon and computation time. This shows the modelling gap that motivated the development of the SCEM model that prioritises speed of prediction over long-term forecast accuracy.

	\begin{sidewaystable*}%[htbp]
		\centering
		\resizebox{\linewidth}{!}{%
		\tabulinesep = 1mm
		\begin{tabu} to 1.5\linewidth {|X[1,c,m]||X[1,c,m]|X[1,c,m]|X[1,c,m]|X[1,c,m]|X[1,c,m]||X[1,c,m]|X[1,c,m]|X[1,c,m]|X[1,c,m]|X[1,c,m]||X[1,c,m]|X[1,c,m]|X[1,c,m]|X[1,c,m]|X[1,c,m]|} \hline
			& \multicolumn{5}{|l||}{Internal Environment Model Complexity}                                                                       & \multicolumn{5}{|l||}{Oil Model Complexity}                                                                                  & \multicolumn{4}{|l||}{Prediction Horizon}                                                                                       &                             \\ \hline
			Oil Model & External data only & Ocean surface + Wind & Wave surface dynamics & Subsurface ocean dynamics & Coupled environment and oil dynamics & Advection + Constant diffusion & Variable diffusion & Mechanical spreading & Subsurface dynamics & Oil weathering & Several hours surface drift & Half day drift & Day drift with oil separation & Long term drift with weathering & Prediction Computation Time \\ \hline \hline
			GNOME     & X             &                      &                       &                           &                                      & X                              &                      & X                      &                       & X                & X          & X                       &                     &                               & Seconds-Minutes                     \\ \hline
			SCEM      & X             & X                    & X                     & X                         &                                      & X                              & X                    & X                      & X                     &                  & X          & X                       & X                   &                               & Seconds-Minutes                     \\ \hline
			MOTHY     & X             & X                    &                       &                           &                                      & X                              & X                    & X                      & X                     & X                & X          & X                       & X                   & X                             & Minutes-Hours               \\ \hline
			MEDSLIK   & X             &                      &                       &                           &                                      & X                              & X                    & X                      & X                     & X                & X          & X                       & X                   & X                             & Minutes-Hours               \\ \hline
			POSEIDON  & X             & X                    & X                     &                           &                                      & X                              & X                    & X                      & X                     & X                & X          & X                       & X                   & X                             & Minutes-Hours               \\ \hline
			sECOM-MDO & X             & X                    & X                     & X                         & X                                    & X                              & X                    & X                      & X                     & X                & X          & X                       & X                   & X                             & Days    \\ \hline                  
		\end{tabu}}
		\caption{\label{tab:oilmodeltable} A qualitative summary of the features and use of several oil models. MEDSLIK, POSEIDON and MOTHY form part of the Mediterranean Decision Support System for Marine Safety and demonstrate their focus on accurate prediction of a spill, but are unsuitable for fast in-the-loop sensor guidance. SCEM displays a use of an internal environment model to supplement external data, the core short-term oil drift dynamics and prediction horizon combined with a short computation time.}
	\end{sidewaystable*}%
	
	The separation of Ocean modelling to Oil modelling does have advantages, allowing for differing hydrodynamic approaches to be used and the appropriation of data from any source, be it small scale Boussinesq models \parencite{Lonin1999}, large scale circulation models \parencite{Marsooli2017} or broad-scale measurements from high-frequency radar, synthetic aperture radar (SAR), wave buoys or other data sources. Furthermore, certain parameters may only need to be calculated where oil is likely to be. Wave spectra for example, could be calculated only where required. However, there are disadvantages in model separation. If the models are not integrated, or run at the same time-steps, large data-sets must be produced and stored by the hydrodynamic model for use by the oil model, which may need to interpolate the data. Also, there can be no two-way coupling between oil and hydrodynamics; the dampening effect of oil on surface waves (integral to SAR measurement) \parencite{Zhang2015a} cannot be included if the hydrodynamics are pre-calculated.
	
	In most cases, the majority of oil volume is contained on the surface, in dark slicks \parencite{ITOPF2011}, with only 10\% in the water column after 24 hours \parencite{Proctor1994}. When subsumed underwater temporarily, depths rarely exceed 10m even in high wind conditions \parencite{Li2013}. This suggests a 2D current simulation, with empirical formula induced variation in depth, a 2D wind simulation and a surface wave model are all sufficient for surface input data into a short-term oil model. Sensitivity studies of a similar model \parencite{DeDominicis2013} demonstrate that a calibrated model retains predictive accuracy for approximately 1-2.5 days, with the forecast accuracy largely dependent upon the input ocean currents. This is sufficient for a prediction horizon of a few hours to a day, for the purpose of model-based sensor guidance.
	
	The novelty of the work herein is the development of a joint model of the hydrodynamics and oil that is computationally efficient for model based guidance. The hydrodynamic model resolves input ocean and wind flow around local bathymetry and geography features too small to be included in the input data. A complete vertical velocity profile is calculated to the sea-bed, using tidal current, Ekman current estimates, Stokes drift and wind induced surface shear. Additionally, a complete wave spectrum is calculated where oil particles are present and environment conditions are contained within each grid cell. Use of spatio-temporally varying external data is also supported if available. The vertical velocity profile is important to estimate the further dispersal of oil resulting from its subsumption and resurfacing within water, without utilising time-intensive 3D flow simulation or large 3D external data-sets that may be unavailable for the local region.
	
	The oil model within SCEM integrates a number of validated algorithms from prior work, with small modifications such as a random walk diffusion correction \parencite{Hunter1993}. The internal hydrodynamic model makes the model suitable for use in regions where high-fidelity external data is absent and can supplement external data by resolving flow around local bathymetry or correcting flow with sensor data. Combined with the contained oil dynamics a complete system is available for analysis and control of sensors, providing a prediction over a short-time horizon (several hours) with prediction error reduced through live sensor information.
	
	\section{Internal Environment Model} \label{environmentmodel}
\begin{figure*}
		\centering
		\includegraphics[width=\textwidth]{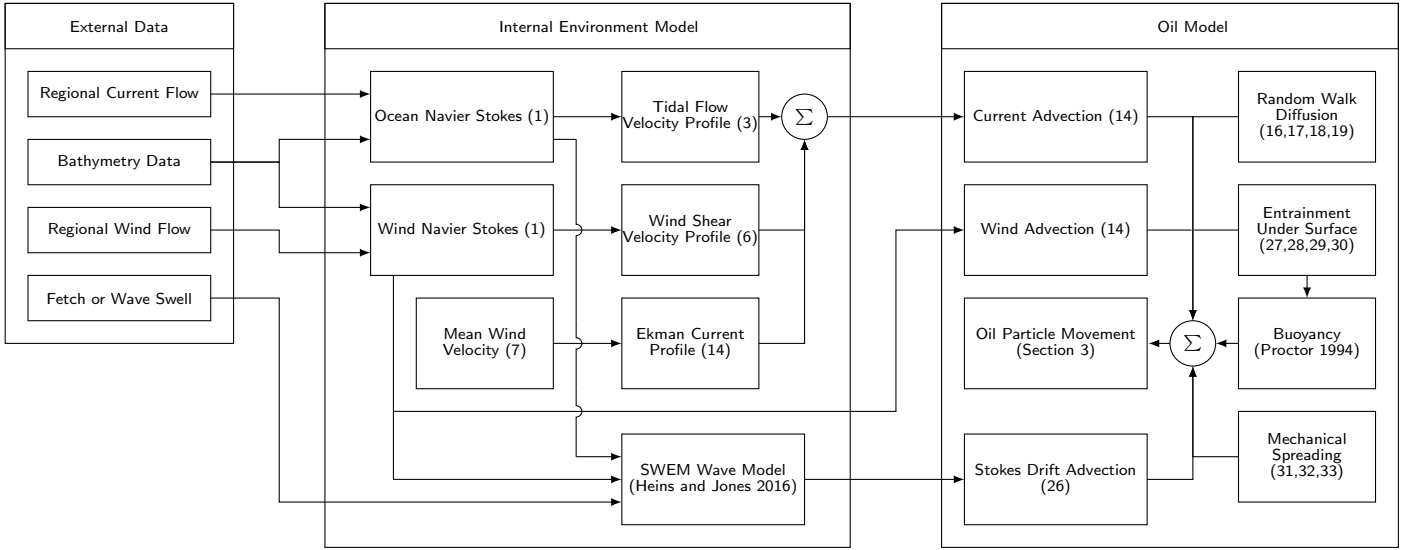}
		\captionsetup{singlelinecheck=off}
		\caption[]{\label{fig:modeldiagram}
			A block diagram of the fluid model, showing the initialisation with external data and the coupling between wind, current and wave motion in producing a contaminant velocity field.
		}
	\end{figure*}

	The environment model contains interconnected sub-systems that describe large scale ocean currents, local currents, local wind and local wave conditions. Figure \ref{fig:modeldiagram} shows the main physics sub-components and their interactions.
	
	 At each time-step the local wind field is calculated first, followed by the large scale ocean current velocity field, then the depth velocity profiles are calculated and finally the wave model is updated to produce a wave induced velocity. These are used, together with oil-only effects such as turbulent diffusion, mechanical spreading, entrainment and buoyancy, to move oil particles. The complete forward simulation algorithm is described in pseudocode in Algorithm \ref{fig:algorithm}.
	
	\begin{algorithm*}[h]
		\SetAlgoLined
		\KwResult{Forwards ocean and contaminant simulation with external/sensor data}
		\tcc{\textbf{INITIALISE\:}}
		read user parameters (domain bounds, empirical parameters, see Table \ref{tab:americagrandeparameters})\;
		load external data files (domain bathymetry, external flow forcing data)\;
		initialise domain\;
		\ForEach{domain grid cell}{
			initialise SWEM wave model\;
			set start date state values\;
		}
		load time-varying contaminant source file (oil type, source location, leak rate)\;
		\tcc{\textbf{RUN SIMULATION\:}}
		\While{start time $\le$ current time $\le$ end time}{
			\tcc{\textbf{CURRENT STATES\:}}
			get predicted state values for current time\;
			get external and/or sensor state values for current time\;
			correct state values using external and/or sensor values at current time\;
			calculate ekman wind value for each grid cell at current time \eqref{eq:ekmanweightedmean}\;
			save state values for current time\;
			
			\BlankLine
			\tcc{\textbf{CURRENT OIL SPILL\:}}
			get oil particles for current time\;
			get external and/or sensor oil values for current time\;
			correct oil spill particles using external and/or sensor values at current time\;
			save oil particles for current time\;
			
			\BlankLine
			\tcc{\textbf{PREDICT NEXT OIL SPILL\:}}
			get corrected state values for current time\;
			calculate time-step\;
			calculate velocity profiles at oil containing grid cells at current time \eqref{eq:tidalvelocityprofile}, \eqref{eq:windshearvelocityprofile}, \eqref{eq:stokesdriftprofile}\;
			simulate SWEM at oil containing grid cells at current time\;
			calculate oil diffusion coefficient at oil containing grid cells at current time \eqref{eq:diffusioncoefficienthorz}\eqref{eq:diffusioncoefficientvert}\;
			calculate total oil velocity profile at oil containing grid cells at current time \eqref{eq:oilmovement}\;
			calculate diffusion correction velocity at oil containing grid cells at current time \eqref{eq:horzdiffcorr}\;
			add source oil particles for time-step\;
			\ForEach{oil particle}{
				\If{oil particle is entrained into water column \eqref{eq:probentrain}}{insert oil particle at calculated depth, set buoyancy to 0 for time-step\;}
				advect oil particle by current time local (total oil velocity + diffusion velocities + correction velocity + buoyancy velocity) for time-step \eqref{eq:oilmovement}\;
				
			}
			calculate oil spill thickness and volume for each grid cell \eqref{eq:oilareacomplex}\;
			\ForEach{oil particle}{
				advect oil particle by local mechanical spreading \eqref{eq:mechanicalspreading}\;
			}
			calculate oil spill thickness and volume for each grid cell \eqref{eq:oilareacomplex}\;
			increase oil particles age\;
			save estimate oil particles for next time\;
			\BlankLine
			\tcc{\textbf{PREDICT NEXT STATES\:}}
			simulate ocean and wind flow for time-step \eqref{eq:navStokes}\;
			save predicted state values for next time\;
			\BlankLine
			\tcc{\textbf{STEP TIME\:}}
			step forward current time by time-step\;
			
		}
		\captionsetup{singlelinecheck=off}
		\caption[]{\label{fig:algorithm}
			Psuedocode of the fluid model, simulating forwards in time.
		}
	\end{algorithm*}
	
	 \subsection{Grid structure}\label{sec:gridstructure}
	 The spatial computation domain is denoted $\Omega \subset \Re^3$, selecting a cuboid section of the Earth including land and ocean with depth. The upper surface $\delta \Omega \subset \Re^2$ of the domain (the water/land to air interface) is discretised upon a regularly spaced grid of $n_x$ grid cells (west to east) and $n_y$ cells (south to north), with equal spacing $\delta x$ and $\delta y$ in the respective directions. Each surface grid-cell represents a Cartesian coordinate volume of $\delta x \delta y \bar{z}_{ij}$, where $[0,\bar{z}_{ij}]$ is the closed interval of water depth in that cell and $\bar{z}_{ij} \coloneqq \bar{z}_{ij}(x_i,y_i) : \delta\Omega \rightarrow \Re_+$ is the average total water depth in that cell area. The set of positive real numbers including 0 is defined by $\Re_+ \subset \Re$. A grid cell at indexed position $(x_i,y_j)$ covers the Cartesian coordinate positions: $(x \pm \frac{\delta x}{2},y \pm \frac{\delta y}{2}, [0,\bar{z}_{ij}])$, where $x_i$ represents the west to east horizontal grid index, $y_j$ is the south to north grid index. 
	 
	 Subsurface water is discretised with a two stage fine and coarse mesh, such that for each grid cell there exists a set of depths $\mathbf{z}(x_i,y_i)$, defined by $\mathbf{z}(x_i,y_i) = \{0,\delta z_1,2 \delta z_1, ... ,N_\text{crit} \delta z_1, z_\text{crit}, z_\text{crit} + \delta z_2, z_\text{crit} + 2 \delta z_2, ... , N_{\bar{z}_{ij}} \delta z_2 \bar{z}_{ij}\}$. Depth spacings $\delta z_1$ and $\delta z_2$ are the finer and coarser vertical grid spacing respectively, $N_\text{crit} \in \mathbb{N}$ is the number of fine mesh grid cells. The switch depth from fine to course mesh, $z_\text{crit}$, is determined by the maximum depth of oil particle insertion into the water column (explained in Section \ref{oilentrainment}), or specified by the user. By utilising a two stage depth grid, finer detail can be maintained near the surface where the majority of contaminant mechanics take place. A 3D grid cell is specified by the indexes $(x_i,y_j,z_w)$, where $z_w$ is the surface to sea floor grid index.

	 \subsubsection{Model States}
	 Each grid-cell is defined by its geo-spatial coordinates and contains the following states:
	 \begin{description}
		\item[$\bullet$] Environmental information (temperature, water density etc).
	 	\item[$\bullet$] Wave spectra.
	 	\item[$\bullet$] Current time wind velocity.
	 	\item[$\bullet$] Previous 12-hour mean wind velocity.
	 	\item[$\bullet$] Tidal flow velocity profile.
	 	\item[$\bullet$] Wind induced surface shear flow velocity profile.
	 	\item[$\bullet$] Ekman current velocity profile.
	 	\item[$\bullet$] Stokes drift velocity profile.
	 \end{description}
 	 The states are formally defined in the following subsections.
	 
	 \subsection{Flow solver} \label{navsolver}
	 A 2D Navier Stokes solver has been implemented to determine local flow velocities for both wind and water, using assumed, measured or external model-provided boundary data. The general form of the Navier Stokes equations, assuming 2-dimensional incompressible flow is:
	 
	 \begin{subequations}\label{eq:navStokes0}
	 	\begin{equation}\label{eq:navStokes}
	 	\frac{\delta \vec{U}}{\delta t} = - (\vec{U} \cdot \nabla)\vec{U} + \nu \nabla^2 \vec{U} - \nabla p + \vec{s}_u,
	 	\end{equation}
	 	\begin{equation}\label{eq:masscons}
	 	\nabla \cdot \vec{U} = 0,
	 	\end{equation}
	 \end{subequations}
	 
	 \noindent
	 where~$\vec{U}(x,y,z,t) : \Omega \times \Re_+ \rightarrow \Re^2$ is the in-plane velocity field such that~$U(x,y,z,t)=[u(x,y,z,t),v(x,y,z,t)]^{\text{T}}$, with $u(x,y,z,t)$ and $v(x,y,z,t)$ the in-plane velocity components in the west to east and south to north directions respectively. For notational brevity the space and time dependency of these variables is not shown in the equations above and similarly throughout the remainder of this paper. In addition~$t \in \Re_+$ is time, $\nu \in \Re$ is the kinematic viscosity of the fluid, $p(x,y,t) : \delta \Omega \times \Re_+ \rightarrow \Re$ is the surface internal pressure field and $\vec{s}_{u} \coloneqq \vec{s}_u(x,y,t) : d\Omega \times \Re_+ \rightarrow \Re^2$ are external surface forces, if present. For wind flow $\vec{U} \coloneqq \vec{U}_{\mathrm{w}} = [u_{\mathrm{w}},v_{\mathrm{w}}]^{\text{T}}$, for current flow $\vec{U} \coloneqq \vec{U}_{\mathrm{c}} = [u_{\mathrm{c}},v_{\mathrm{c}}]^{\text{T}}$ and for Ekman wind $\vec{U} \coloneqq \vec{U}_{\mathrm{E}} = [u_{\mathrm{E}},v_{\mathrm{E}}]^{\text{T}}$.  Flow is calculated for ocean surface currents and for wind velocities at 10m above sea level by solving~\eqref{eq:navStokes0}~ subject to initial conditions on velocity at the simulation start time. These are set from external data, or by setting $\vec{U}(x,y,z,0)$ to a best-estimate of mean flow if no data is available. Boundary conditions are described in the next section.
	 
	The Navier-Stokes equations are spatially discretised upon a staggered grid \parencite{F.Harlow1965}, with spatial derivatives approximated by finite differences. With respect to time-stepping, diffusion terms are solved using a backward Euler method and Gauss Seidel Successive Over Relaxation \parencite{Stam2001}, whilst advective terms are solved using a forward Euler method. Mass conservation is enforced via an iterative pressure projection step, in which the pressure field is found using Gauss-Seidel Successive Over Relaxation~\parencite{Stam2001}, with subsequent correction of the velocity field. This is repeated until the flow-field divergence is below a nominal tolerance. The time-step is variable with the step size determined by the Courant number~\parencite{Courant}
	 
	 \subsubsection{Boundaries, measurements and obstacles}
	 Obstacles are regions of $\vec{U}_{\mathrm{c}} = 0$ for ocean current flow velocity, or $\vec{U}_{\mathrm{w}} : \; \norm{\vec{U}_{\mathrm{w}}}_2 \le \kappa^2 \norm{\vec{U}_{\mathrm{w}_{\text{max}}}}_2$ for wind flow where $\kappa \coloneqq \kappa(x,y) : \delta \Omega \rightarrow \Re_+$ is a wind resistance coefficient based on the environment and $\vec{U}_{\mathrm{w}_{\text{max}}}$ the maximum wind velocity. The presence of obstacles, such as coastline geography, is accounted for by the use of Dirichlet boundary conditions on the velocity field in relevant grid cells. Due to the staggered grid implementation, this is a form of semi-slip boundary \parencite{F.Harlow1965}. This is not unprecedented in ocean models, a user selected value for slip is found in the NEMO ocean model \parencite{Madec2011}, with large scale models using free-slip and small-scale models using no-slip. A semi-slip induces the circulation expected from boundary layers but avoids under-estimation of fluid velocities in sparse grids near walls.
	 
	 Velocity field information from measurements or external data can either be set precisely or within a bounded range, between a minimum and maximum value, representing the accuracy of the sensor. The measured value of an uncertain measurement is applied prior to the projection step of flow calculation. During pressure projection the value is altered, within the bounded range, to ensure divergence free flow. If the value is at a boundary limit, or is assumed accurate (a recent measurement), then it is fixed during pressure projection and other free flow-field velocities are adjusted by the pressure field until the flow is divergence free.
	 
	 Domain edge boundary conditions can be specified as Dirichlet conditions on velocity, or left open as free-flow.
	 
	 \subsection{Wind flow}
	 Calculation of wind velocity $\vec{U}_{\mathrm{w}}$ is handled by the flow solver described in Section \ref{navsolver}, with the replacement of zero flow boundaries for obstacles by maximum wind-speed conditions  to represent wind resistant areas such as cities or mountains. This acts as a flow restriction and thus resolves flow to greater accuracy for local geographic features. The velocity limit $\vec{U}_{\mathrm{w}_\text{lim}} \coloneqq \vec{U}_{\mathrm{w}_\text{lim}}(x,y) : \delta \Omega \rightarrow \Re^2$ is calculated by the urban canopy profile
	 
	 \begin{equation}
	 \vec{U}_{\mathrm{w}_\text{lim}} = (1-\lambda_{p})^{2}\norm{\vec{U}_{\mathrm{w}_{\text{max}}}}_2,
	 \end{equation}
	 	 
	 \noindent
	 where $\lambda_{p} \coloneqq \lambda_{p}(x,y) : \delta \Omega \rightarrow [0,1] \subset \Re$ is the obstruction plan, or footprint, density in the cell area at 10m altitude \parencite{CERC2017}. If an obstruction density map is not available, a coefficient can be specified in place of $(1-\lambda_{p})^{2}$ for all coastal and land cells.
	 
	 \subsection{Ocean flow}
\begin{figure}[H]
	 	\centering
	 	\includegraphics[width=0.48\textwidth]{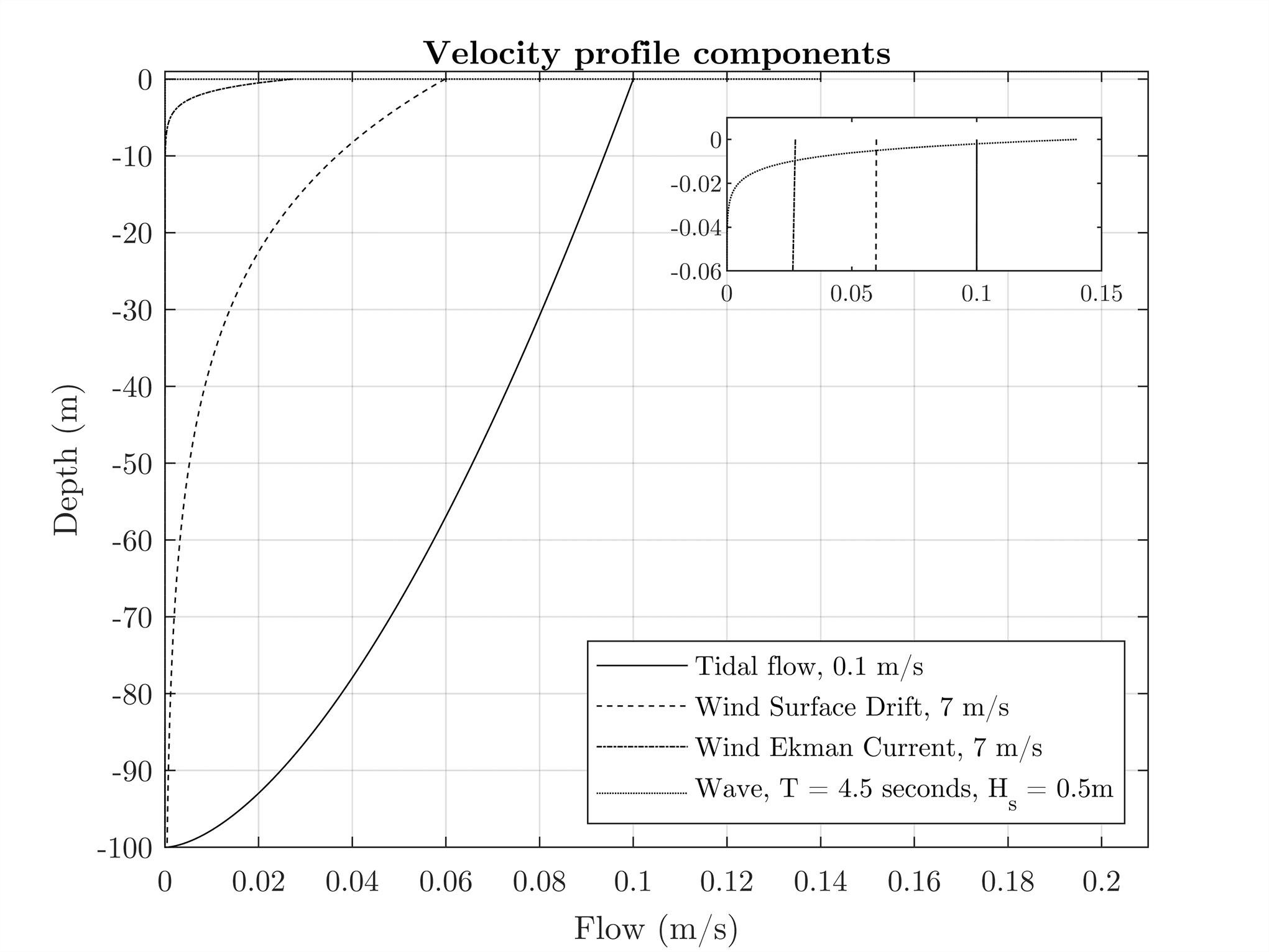}
	 	\caption{\label{fig:depthflowmechanisms} A depiction of sub-surface flow resulting from mechanisms included in the three-dimensional model. The insert magnifies the sub-surface flow at shallow depths, note the very shallow effect of wind surface drift.}
	 \end{figure}

	 The Navier-Stokes equation are solved in 2D, but a depth velocity profile is important in calculating oil-trajectories and inducing the separation of slicks caused by sub-surface shear flows and entrainment. Typical individual velocity profiles are shown in Figure \ref{fig:depthflowmechanisms}.
	 
	 \subsubsection{Tidal and circulation flow}
	 The velocity profile, introducing vertical variation to $\vec{U}_{\mathrm{c}}$ for a tide driven flow, follows a standard logarithmic profile where $z \in [0,\bar{z}] \subset \Re$ is depth in the water column, with $z=0$ at the surface and $z=\bar{z}$ at maximum water depth where $\bar{z}$ is the mean total water depth in that cell area as in section \ref{sec:gridstructure}. A no-slip condition, $\vec{U}_{\mathrm{c}_{\bar{z}}} = 0$ is imposed on the sea floor and $\vec{U}_{\mathrm{c}_z}$ increases to its maximum value at the surface.  As predictions focus on surface oil particles, boundary layer simulation is omitted and $U_{\mathrm{c}_z}$ can be simply defined as \parencite{Thiebaut2016}
	 
	 \begin{equation}\label{eq:tidalvelocityprofile}
	 \vec{U}_{\mathrm{c}_z} = \vec{U}_{\mathrm{c}_0} \left(1-\frac{z}{\bar{z}}\right)^{\frac{1}{6}},
	 \end{equation}
	 \noindent
	 where the empirical denominator parameter in the power law has been assigned the value of 6, which falls in the range of accepted values for ebbing and flowing tides.
	 
	 \subsubsection{Wind induced surface shear}
	 
	 Under strong wind conditions the velocity of surface water is heavily affected by the boundary stress between the two-phase flow of air and water, so is vital for inclusion in an advective ocean model. Large scale models often use measured wind speed data \parencite{DeDominicis2013a} or wind speed estimated from surface roughness (measured via radar scattering) \parencite{Smith1988} to simply calculate a surface flow velocity. This takes the form of a scaled velocity $\alpha_{\mathrm{w}} \norm{\vec{U}_{\mathrm{w}}}_2$, rotated by a wind drift angle $\beta \coloneqq \beta(x,y,t) : \delta \Omega \times \Re_+ \rightarrow \Re$, representing the balancing of wind shear and coriolis effects. This velocity is then scaled to a logarithmic velocity profile \parencite{Wu1975}, modelled in oceans as beginning at $z_0$ (the wind driven surface layer) and falling to zero effect at $z_c$ meters \parencite{Proctor1994}. This latter depth can be approximated as
	 
	 \begin{equation} \label{eq:proctorz}
	 z_c \approx \alpha_z L.
	 \end{equation}
	 
	 A value of $\alpha_z = 2$ is suggested to give good agreement with observations \parencite{Elliott1986} in a short-fetch environment, using $L$ as the dominant wavelength of sea-surface waves. For even a low wind speed fetch in deep water, wave lengths are likely to be around 8 meters giving rise to large $z_c$ values and a large effect of wind on sub-surface currents, with data supporting a wind penetration depth of 40m \parencite{Elliott1986}.
	 
	 An assumption of the above method is instantaneous changes of sub-surface currents in response to local wind gusting. Here, wind effects are modelled in parts, deep effects are modelled as a combination of slow time-varying Ekman currents and stokes drift from a linear wave model. Shallow effects are instantaneously applied by a logarithmic velocity profile. In a wind wave spectrum, local wind affects only the small-scale ripples (capillary waves) and gravity-wind waves that are accounted for in the linear wave model. Modelling just capillary waves gives rise to varying surface roughness (as wave amplitude) across wind conditions, of typical wavelength \parencite{Lamb1895} defined by
	 
	 \begin{equation}\label{eq:capillarywavelength}
	 L_{\text{capillary}} = 2 \pi \sqrt{\frac{\sigma_{\text{water}}}{(\rho_{\text{water}} - \rho_{\text{air}})g}},
	 \end{equation}
	 \noindent
	 where for an air-water interface, $L_{\text{capillary}} = 0.017$m \parencite{Lamb1895} and $\sigma_{\text{water}}$ is the surface tension of water, $\rho_{\text{water}}$ and $\rho_{\text{air}}$ are the densities of water and air, respectively. Thus the new wind shear zero effect depth for~$\alpha_z=2$ is 0.037m when using \eqref{eq:capillarywavelength} to determine the wavelength in \ref{eq:proctorz}. This is a shallow depth, where viscous shear and vertical mixing allows an assumption of a velocity change time-scale much smaller than the simulation time-step. Hence velocity changes immediately with fluctuating wind as in traditional percentage based algorithms for surface oil spill drift due to wind/wave interaction \parencite{Spaulding2017}, with the wind shear velocity profile $\vec{U}_{\mathrm{w}_z} \coloneqq \vec{U}_{\mathrm{w}_z}(x,y,z,t) : \Omega \times \Re_+ \rightarrow \Re^2$, defined by
	 
	 \begin{equation}\label{eq:windshearvelocityprofile}
	 \vec{U}_{\mathrm{w}_z} = \alpha_{\mathrm{w}} \vec{U}_{\mathrm{w}} e^{- \frac{2 \pi}{z_c} z},
	 \end{equation}
	 \noindent
	 where $\alpha_{\mathrm{w}} \in [0.005,0.03] \subset \Re$. A value of $0.02$ is suggested for $\alpha_{\mathrm{w}}$ \parencite{Proctor1994}, but varies within literature \parencite{Kim2014}.
	 
	 \subsubsection{Ekman currents}
	 Ekman currents describe the net motion of fluid that results from the balance of a forcing wind, turbulent drag and Coriolis forces. In a small scale simulation it would be preferable not to assume an instantaneous (in distance and time) change in the sub-surface layer velocity due to wind. Ekman currents take approximately 12 hours to form ($T_{\mathrm{E}} = 12$ hours), accelerating approximately linearly to their fully formed magnitude \parencite{Weatherly1975}. Ideally, the Ekman current would change towards its final value at each time-step, but this would require changing every depth value in every grid cell, at every time-step, leading to excessive computational load. An alternative would be to keep a moving average of the last 12 hours of wind data, but this requires stored data and introduces a large phase lag in Ekman changes. For a domain where wind-speed changes are frequent, an incremental weighted mean of wind speeds to form an average of the past 12 hours of wind speed is proposed. The Ekman wind velocity $\vec{U}_{\mathrm{w}_{\mathrm{E}}} \coloneqq \vec{U}_{\mathrm{w}_{\mathrm{E}}}(x,y,t) : \delta \Omega \times \Re_+ \rightarrow \Re^2$, is calculated for the surface of each grid cell:
	 
	 \begin{equation}\label{eq:ekmanweightedmean}
	 \vec{U}_{\mathrm{w}_{\mathrm{E}}} = \frac{W_{\mathrm{E}_1} \vec{U}_{\mathrm{w}_{\mathrm{E}}}^{t - \delta t} + W_{\mathrm{E}_2} \vec{U}_{\mathrm{w}}^t} {W_{\mathrm{E}_1} + W_{\mathrm{E}_2}},
	 \end{equation}
	 \noindent
	 where the Ekman averaged wind velocity at the previous time-step is $\vec{U}_{\mathrm{w}_{\mathrm{E}}}^{t - \delta t}$ and $\vec{U}_{\mathrm{w}}^t$ is the current time wind-velocity. The weights for the value $W_{\mathrm{E}_1} \in \Re : \; 0 \le W_{\mathrm{E}_1} \le 1$ and future values $W_{\mathrm{E}_2} \in \Re : \; 0 \le W_{\mathrm{E}_2} \le 1$ are calculated as
	 \begin{subequations}
	 \begin{equation}
	 W_{\mathrm{E}_1} = \frac{T_{\mathrm{E}} - \delta t}{\frac{1}{2} T_{\mathrm{E}}}
	 \end{equation}
	 and
	 \begin{equation}
	 W_{\mathrm{E}_2} = \frac{\delta t}{T_{\mathrm{E}}}.
	 \end{equation}
	\end{subequations}
	 
\begin{figure*}
	 	\centering
	 	\makebox[\textwidth][c]{\includegraphics[width=\textwidth]{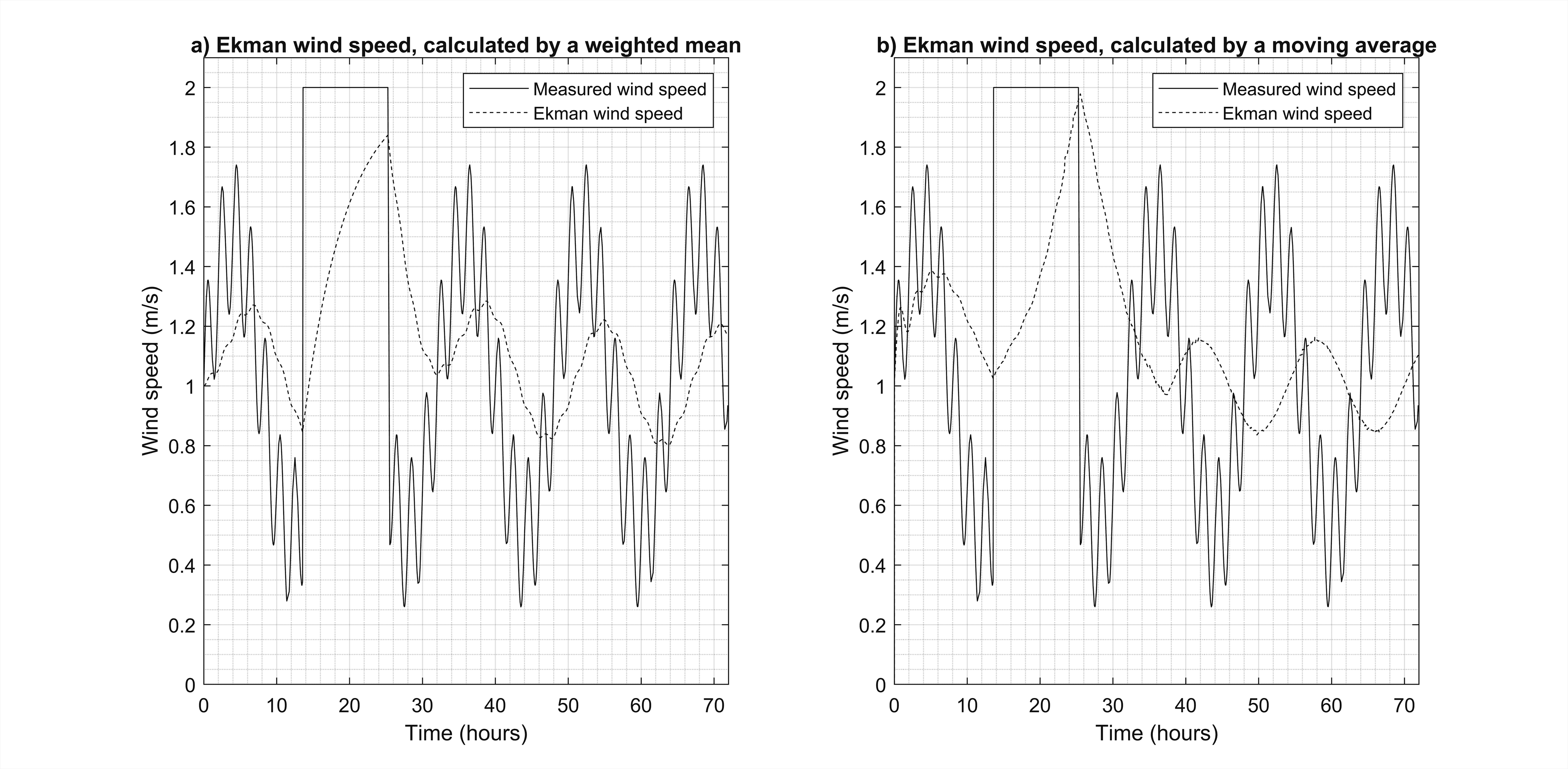}}
	 	\caption{\label{fig:ekmanwithtime} A depiction of the wind speed $\vec{U}_{\mathrm{w}}$ that Ekman currents are calculated from, $\vec{U}_{\mathrm{w}_{\mathrm{E}}}$, under noisy wind conditions. a) Calculated using the weighted mean approach. b) Calculated by a traditional 12 hour moving average approach, which shows a more linear growth but significantly greater lag.}
	 \end{figure*}

	  Figure \ref{fig:ekmanwithtime} shows the growth and decay of the Ekman wind speed, used to calculate the Ekman current, under a range of wind conditions when calculated by both the weighted mean and moving average approaches. Results from literature suggest there should be no lag between wind stress and Ekman shelf velocities~\parencite{Kirincich2008}. Whilst both approaches induce a lag in the Ekman current velocity, inspection of~Figure~\ref{fig:ekmanwithtime} clearly shows the lag from the moving average approach is significantly greater than that from the weighted mean approach, to the point where the Ekman wind speed response is almost completely out of phase with the forcing wind.
	 
	 The wind stress of the surface layer can be used to estimate the Ekman current magnitude. Prior work provides stress coefficients for water under a variety of conditions, including adjustment factors for wind speeds measured at various heights to normalise their values at 10m above sea level \parencite{Wu1980} and \parencite{Smith1988}. Let the stress coefficient $C_{D_{\text{stress}}} \coloneqq C_{D_{\text{stress}}}(x,y,t,\norm{\vec{U}_{\mathrm{w}_{\mathrm{E}}}}_2) : \delta \Omega \times \Re_+ \times \Re \rightarrow \Re_+$. Hence, using $\vec{U}_{\mathrm{w}_{\mathrm{E}}}$ as wind velocity yields
	 
	 \begin{equation}
	 \tau = C_{D_{\text{stress}}} \rho_{\text{air}} \norm{\vec{U}_{\mathrm{w}_{\mathrm{E}}}}_{2}^2,
	 \end{equation}
	 \noindent
	 where the scalars $\tau \coloneqq \tau(x,y,t) : \delta \Omega \times \Re_+ \in \Re_+$ are the wind stresses on the water surface and $\rho_{\text{air}} \in \Re_+$ is the air density. A final Ekman current velocity profile is calculated \parencite{pond2013introductory}, using a vertical eddy viscosity coefficient \parencite{Rasmussen1985} of $A_{z} \coloneqq A_{z}(x,y,t) : \delta \Omega \times \Re_+ \rightarrow \Re_+$  and a surface Ekman speed $V_{0_{\mathrm{E}}} \coloneqq V_{0_{\mathrm{E}}}(x,y,t) : \delta \Omega \times \Re_+ \rightarrow \Re_+$ defined by
	 
	 \begin{subequations}
	 \begin{equation}
	 A_{z} = 4.3 \times 10^{-4} \norm{\vec{U}_{\mathrm{w}_{\mathrm{E}}}}_{2}^2
	 \end{equation}
	 and
	 \begin{equation}
	 V_{0_{\mathrm{E}}} = \frac{\sqrt{2} \pi \tau}{z_{\mathrm{E}} \rho_{\text{water}} |f|},
	 \end{equation}
	 \end{subequations}
	 \noindent
	 where $z_{\mathrm{E}} \coloneqq z_{\mathrm{E}}(x,y,z,t) : \Omega \times \Re_+ \rightarrow \Re_+$ is the Ekman layer depth \parencite{pond2013introductory}, $f \coloneqq f(x,y,t) : \delta \Omega \times \Re_+ \rightarrow \Re$ is the coriolis frequency and $\rho_{\text{water}} \in \Re_+$ is the water density. Adjusting for a coordinate system where $u$ is positive east velocity and $v$ is positive north, with an ascending $z$ with depth and positive clockwise from north angles, an alternative formulation that also reflects the smaller drift divergence angle in current formations under high wind conditions can be described by
\begin{subequations}	 \label{eq:ekmanvelocityprofile}
	 \begin{equation}\label{eq:ekmanuvelocityprofile}
	 u_{\mathrm{E}_z} = \pm V_{0_{\mathrm{E}}} \sin \left( \beta_{\text{rad}} - \frac{\pi}{z_{\mathrm{E}}} z \right) e^{-\frac{\pi}{z_{\mathrm{E}}} z},
	 \end{equation}
	 \noindent
	 where the negative sign applies to the northern hemisphere, the positive to the southern hemisphere. Similarly 
	 
	 \begin{equation}\label{eq:ekmanvvelocityprofile}
	 v_{\mathrm{E}_z} = \pm V_{0_{\mathrm{E}}} \cos \left( \beta_{\text{rad}} - \frac{\pi}{z_{\mathrm{E}}} z \right) e^{-\frac{\pi}{z_{\mathrm{E}}} z},
	 \end{equation}
\end{subequations}
	 \noindent
	 where a wind drift angle \parencite{Wang2010} is proposed for the Ekman current angle, instead of a constant 45 degrees:
	 
	 \begin{subequations}
	 \begin{equation}
	 \beta = \begin{cases}40^{ \circ  }-8^{ \circ  }\sqrt [ 4 ]{ u_{\mathrm{w}}^{ 2 }+v_{\mathrm{w}}^{ 2 } } \hspace{0.5em} &\text{for} \hspace{0.5em} 0 \hspace{0.5em} \le \sqrt { u_{\mathrm{w}}^{ 2 }+v_{\mathrm{w}}^{ 2 } } \le \hspace{0.5em} 25~\!\textrm{m/s} \\ 
	 0^{ \circ  } &\text{for} \hspace{0.5em} \sqrt { u_{\mathrm{w}}^{ 2 }+v_{\mathrm{w}}^{ 2 } } > \hspace{0.5em} 25~\!\textrm{m/s} \end{cases},
	 \end{equation}
	 and
	 \begin{equation}
	 \beta_{\text{rad}} = \beta \frac{\pi}{180}.
	 \end{equation}
	 \end{subequations}
	 
	 The velocity components~\eqref{eq:ekmanvelocityprofile} compose the fully formed Ekman velocity~$\vec{U}_{\mathrm{c}_{\mathrm{E}_z}} = [u_{\mathrm{E}_z},v_{\mathrm{E}_z}]^{T}$, where~$\vec{U}_{\mathrm{c}_{\mathrm{E}_z}} \coloneqq \vec{U}_{\mathrm{c}_{\mathrm{E}_z}}(x,y,z,t) : \Omega \times \Re_+ \rightarrow \Re^2$.
	 
\begin{figure}[H]
	 	\centering
	 	\includegraphics[width=0.48\textwidth]{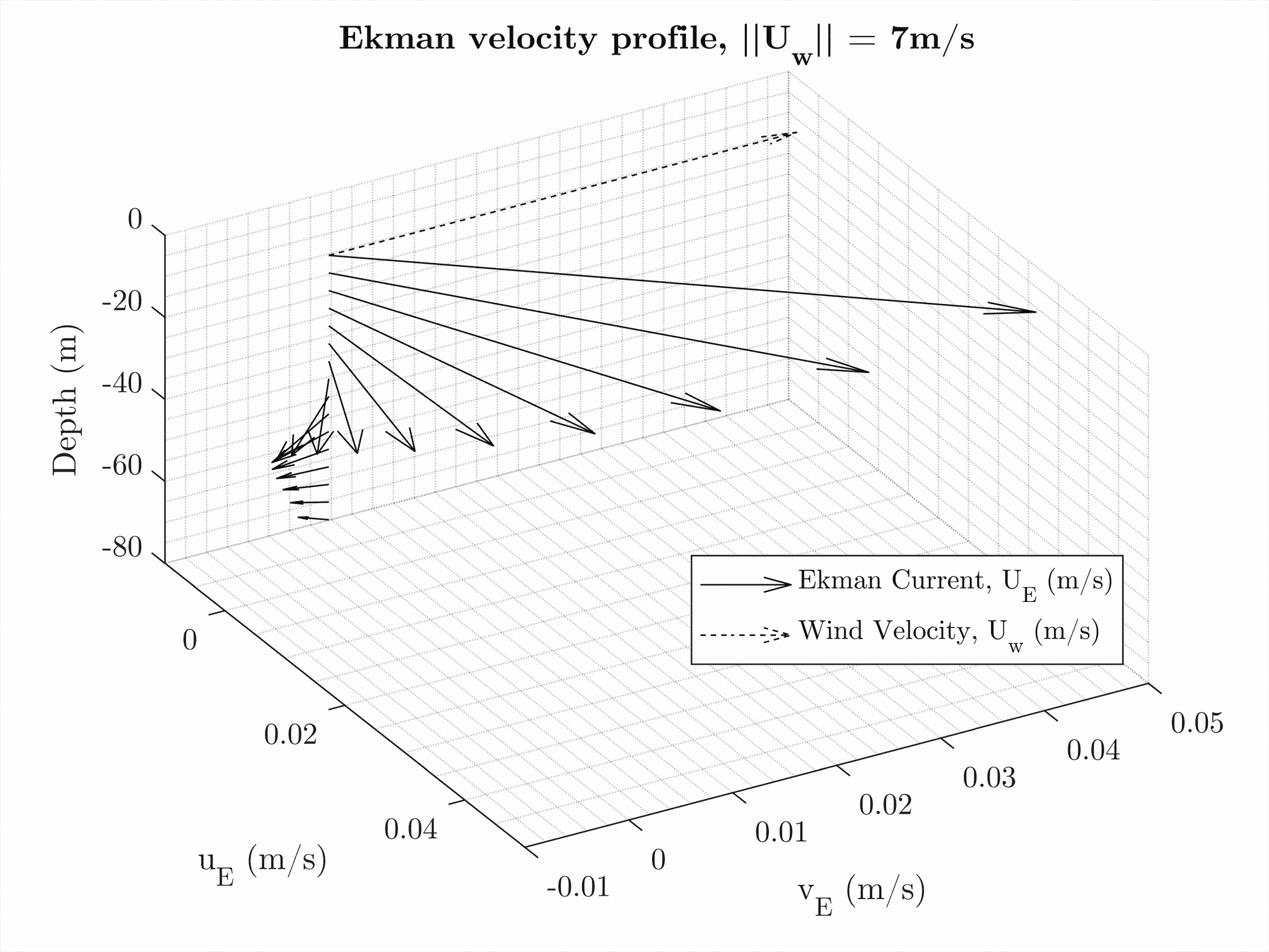}
	 	\caption{\label{fig:watervelwithdepth} The spatial variation of Ekman velocity with depth resulting from non-aligned wind and current angles.}
	 \end{figure}

	 These equations produce an Ekman velocity profile, shown in Figure \ref{fig:watervelwithdepth}, that follows a typical spiral pattern and has a magnitude of approximately 1\% of the wind speed. This is as expected, the 3\% wind velocity advection employed by classical models will be a composite of the smaller Ekman currents, Stokes drift and Surface stress induced currents calculated separately here.
	 
	 \subsection{Linear wave model}
	  To determine the effect of waves on contaminants a spatio-temporally varying wave spectrum is approximated by the Sheffield Wave Environment Model (SWEM) \parencite{Heins}, which combines modified wave spectra from ocean swell, local wind, surface current and finite water depth to simulate the ocean surface. It includes a directional spreading function and swell estimation from fetch parameters or buoy data. Each cell updates its wave model with the local wind and surface current velocity at every time-step. The wave models then re-evaluate the wave spectra, along with the significant wave height $H_{s} \coloneqq H_{s}(x,y,t) : \delta \Omega \times \Re_+ \rightarrow \Re_+$, wavelength $L \coloneqq L(x,y,t) : \delta \Omega \times \Re_+ \rightarrow \Re_+$ and wave period $T \coloneqq T(x,y,t) : \delta \Omega \times \Re_+ \rightarrow \Re_+$ for each grid cell. In a time-constrained simulation, the wave model can be updated only in cells where oil is present without adversely affecting results.
	
	\section{Oil model} \label{oilmodel}
	The oil model uses a common Lagrangian approach \parencite{Spaulding2017}, utilising large numbers of particles (see section \ref{sec:numberofoilparticles}), each representing a volume of contaminant. Particles undergo advection and turbulent diffusion in response to forcing from the environmental model. Particles are then used to build a thickness map and undergo mechanical spreading in areas where the thickness is above a minimum value, with particle size determined from oil properties. Particles can be entrained underwater, determined by variables from the wave model, with subsequent resurfacing dependent upon vertical turbulent diffusion and terminal buoyancy velocity.
	
	\subsection{Advection and diffusion} \label{oiladvectionanddiffusion}
	The advective velocity of particles at depth $z$ consists of horizontal velocity components $u_{\mathrm{o}_z}$, $v_{\mathrm{o}_z}$ and a vertical velocity component $w_{\mathrm{o}_z}$. These are determined from a summation of tidal, wind induced surface shear and Ekman current velocities, plus turbulent diffusion terms as follows:
	\begin{equation}\label{eq:oilmovement}
	\begin{aligned}
	\begin{bmatrix} u_{\mathrm{o}_z} \\ v_{\mathrm{o}_z} \\ w_{\mathrm{o}_z} \end{bmatrix} &=& \alpha_{\mathrm{w}_{\mathrm{o}}}\begin{bmatrix} u_{\mathrm{w}} \\ v_{\mathrm{w}} \\ 0 \end{bmatrix} &+&\alpha_{\mathrm{c}_{\mathrm{o}}}\begin{bmatrix} u_{ c_z }\\ v_{ c_z } \\ 0 \end{bmatrix} &&\\	
	&+& \begin{bmatrix} u_{\mathrm{w}_z} \\ v_{\mathrm{w}_z} \\ 0 \end{bmatrix} &+& \begin{bmatrix} u_{{E_z}}\\ v_{{E_z}} \\ 0 \end{bmatrix} &+& \begin{bmatrix} u_{{s_z}}\\ v_{{s_z}} \\ 0 \end{bmatrix} \\	
	&+& \begin{bmatrix} u_d \\ v_d \\  0 \end{bmatrix} &+& \begin{bmatrix} u' \\ v' \\  w_z' \end{bmatrix} &&,
	\end{aligned}
	\end{equation}
	\noindent
	where $\alpha_{\mathrm{w}_{\mathrm{o}}} \in \Re : \; 0 \le \alpha_{\mathrm{w}_{\mathrm{o}}} \le 0.05$ is a coefficient for additional wind advection and $\alpha_{\mathrm{c}_{\mathrm{o}}}\in \Re : \; 0.95 \le \alpha_{\mathrm{c}_{\mathrm{o}}} \le 1.1$ is an advection coefficient for tidal currents. The diffusion correction velocities $u_d \coloneqq u_d(x,y,t,D_h)  : \delta \Omega \times \Re_+ \times \Re \rightarrow \Re$ and $v_d \coloneqq v_d(x,y,t,D_h) : \delta \Omega \times \Re_+ \times \Re \rightarrow \Re$ are defined 
	\begin{subequations}\label{eq:horzdiffcorr}
		\begin{equation}
		u_d(x,y,t,D_h) = \frac{\delta D_h}{\delta x},
		\end{equation}
		in the horizontal $x$ direction and
		\begin{equation}
		v_d(x,y,t,D_h) = \frac{\delta D_h}{\delta y}
		\end{equation}
		in the horizontal $y$ direction.
	\end{subequations}
	They are the spatial derivative of $D_{h} \coloneqq D_{h}(x,y,t) : \delta \Omega \times \Re_+ \rightarrow \Re$, the horizontal diffusion coefficient \parencite{Hunter1993}. The turbulent diffusion velocities comprise of $u' \coloneqq u'(x,y,t) : \delta \Omega \times \Re_+ \in \Re$, $v' \coloneqq v'(x,y,t) : \delta \Omega \times \Re_+ \in \Re$, $w' \coloneqq w'(x,y,z,t) : \Omega \times \Re_+ \in \Re$ in the horizontal $x$ and $y$ plane and vertical $z$ direction respectively. The stokes drift velocities (defined in Section \ref{stokesdrift}) $u_{sz} \coloneqq u_{sz}(x,y,z,t) : \Omega \times \Re_+ \in \Re$, $v_{sz} \coloneqq v_{sz}(x,y,z,t) : \Omega \times \Re_+ \in \Re$ in the $x$ and $y$ horizontal direction respectively. The additional wind advection represents only the carrying of oil droplets by wind, since the major wind drift is accounted for in the hydrodynamic model.
	
	Turbulent diffusion is calculated by the common random walk method \parencite{Spaulding2017}, but avoids direct parameter setting for horizontal diffusivity and vertical diffusivity coefficients in favour of empirical formulae that also introduce variation in the diffusion coefficient dependent upon flow properties. Spatial variation in diffusion coefficient results in a requirement for a diffusion correction velocity \parencite{Hunter1993}. Horizontal turbulent diffusion velocity is assumed constant with depth and calculated \parencite{Chao2003} using:
\begin{subequations}	
	\begin{equation}
	u' = \xi \sqrt{\frac{12 D_{h}}{\delta t}} \sin(2 \pi \phi),
	\end{equation}
	
	\begin{equation}
	v' = \xi \sqrt{\frac{12 D_{h}}{\delta t}} \cos(2 \pi \phi),
	\end{equation}
\end{subequations}
	\noindent
	where $\xi$ and $\phi$ are particle specific random variables with uniform distribution in $[0,1]$. Vertical turbulent diffusion velocity is depth dependent and calculated \parencite{Lardner2000} according to:
	
	\begin{equation}
	w' = (2 \zeta - 1) \sqrt{\frac{6 D_{v_z}}{\delta t}},
	\end{equation}
	\noindent
	where $\zeta$ is a particle specific random variable with uniform distribution in $[0,1]$. Coefficients for horizontal diffusivity $D_{h}$ \parencite{Baldauf2012} and vertical diffusivity $D_{v_z} \coloneqq D_{v_z}(x,y,t) : \delta \Omega \times \Re_+ \rightarrow \Re_+$ \parencite{Li2013} are calculated as follows:
	
	\begin{subequations}
		\begin{equation}\label{eq:diffusioncoefficienthorz}
		D_{h} = \frac{c_{\text{smag}}}{\frac{1}{\delta x^2 + \delta y^2}} \sqrt{T_{\text{smag}} + S_{\text{smag}}},
		\end{equation}
		
		\begin{equation}
		T_{\text{smag}} = \frac{\delta u_{\mathrm{c}}}{\delta x} - \frac{\delta v_{\mathrm{c}}}{\delta y},
		\end{equation}
		
		\begin{equation}
		S_{\text{smag}} = \frac{\delta u_{\mathrm{c}}}{\delta y} + \frac{\delta v_{\mathrm{c}}}{\delta x},
		\end{equation}
		
		\begin{equation}\label{eq:diffusioncoefficientvert}
		D_{v_z} = 0.028 \frac{H_{s}^{2}}{T} e^{-2 \frac{1}{L} z},
		\end{equation}
	\end{subequations}
	\noindent
	where $c_{\text{smag}} \in \Re : \; 0.01 \le z \le 0.3$ is an empirical coefficient, with a nominal default value of $0.1$. 
	
	\subsubsection{Stokes Drift}\label{stokesdrift}
	Stokes drift is the net horizontal movement of a particle due to wave motion, resulting from shear stresses and mixing layers from surface gravity waves. For each grid cell in which there are oil particles and for each time-step, the spectral wave model SWEM is used to compute the wave parameters that govern~Stokes drift, chiefly significant wave height~$H_\mathrm{s}$, wavelength~$L$ and wave period ~$T$. These are evaluated from the peak magnitude $a_\mathrm{p} \coloneqq a_\mathrm{p}(x,y,t) : \delta \Omega \times \Re_+ \rightarrow \Re_+$ and corresponding peak frequency $f_\mathrm{p} \coloneqq f_\mathrm{p}(x,y,t) : \delta \Omega \times \Re_+ \rightarrow \Re_+$ of the wave spectrum.
	
	Webb proposes the use of the peak frequency, with a Stokes drift amplitude modified by the spectral moment (calculated through intergrands) and empirical terms specific to that spectrum \parencite{Webb2011}. SWEM's spectrum is a summation of several others and therefore this approach would require multiple calculations of spectral moments and ultimately, too much computation. Hence, only the peak information of the SWEM spectrum (representing the fetch, local current and local wind interaction) is used, as the high frequency ripple waves are accounted for through wind shear.
	
	Stokes drift magnitude is similar to a near-surface tidal shear \parencite{Elliott1986} or 1 - 2$\%$ of the wind speed \parencite{Proctor1994}. The literature suggested \parencite{Elliott1986} hyperbolic trigonometric formulation of Stokes drift can become undefined in deep water conditions, hence it is redefined to give the Stokes drift speed:
	
	\begin{equation}
	\norm{\vec{U}_{{s}_z}}_2 = \omega k a_{p}^{2} e^{-2 k z},
	\end{equation}
	\noindent
	where $\omega = 2 \pi / T_{\text{peak}}$, $k  = 2\pi / L_{\text{peak}}$ using the wave spectrum peak values from SWEM. To achieve an accurate Stokes drift velocity, the wave spectrum produces an average wave energy direction and scales the Stokes drift velocity to the proportion of wave energy in that direction compared to the total wave energy in the spectrum. The direction and magnitudes of the waves are expressed in polar coordinates as follows:
\begin{subequations}	
	\begin{equation}
	\theta_{{\Psi_{\text{Total}}}_i} = \atantwo \left( \frac{k_y}{k_x} \right),
	\end{equation}
	
	\begin{equation}
	r_{{\Psi_{\text{Total}}}_i} = \Psi_{\text{Total}}(k_x,k_y),
	\end{equation}
\end{subequations}
	\noindent
	where $\Psi_{\text{Total}}(k_x,k_y) \coloneqq \Psi_{\text{Total}}(k_x(x,y,t),k_y(x,y,t)) : \delta \Omega \times \Re_+ \rightarrow \Re_+$ is the energy of the waves with wavenumbers $k_x \coloneqq k_x(x,y,t) : \delta \Omega \times \Re_+ \rightarrow \Re$ and $k_y \coloneqq k_y(x,y,t) : \delta \Omega \times \Re_+ \rightarrow \Re$. The polar angle $\theta_{{\Psi_{\text{Total}}}_i} \coloneqq \theta_{{\Psi_{\text{Total}}}_i}(x,y,t) : \delta \Omega \times \Re_+ \rightarrow \Re$ and magnitude $r_{{\Psi_{\text{Total}}}_i} \coloneqq r_{{\Psi_{\text{Total}}}_i}(x,y,t) : \delta \Omega \times \Re_+ \rightarrow \Re_+$ form the polar coordinate representation of that wave-number, with magnitude being the wave energy and angle as the wave direction. The wave spectrum is thus converted from 2D $[k_x,k_y]$ wave numbers to a $k_x k_y$ by 1 vector of polar coordinates. The sum of the vector of polar coordinates provides an average wave energy polar coordinate with magnitude and direction of the average wave energy:
\begin{subequations}	
\textcolor{red}{}
	\begin{equation}
	[\theta_{\text{sum}},r_{\text{sum}}] = \sum _{ i=1 }^{ k_x k_y }{ [\theta_{{\Psi_{\text{Total}}}_i}, r_{{\Psi_{\text{Total}}}_i}  ]} ,
	\end{equation}
	
	\begin{equation}
	\Psi_{\text{avg}_{\theta}} = r_{\text{sum}}.
	\end{equation}
\end{subequations}	
	The polar angle $\theta_{\text{sum}} \coloneqq \theta_{\text{sum}}(x,y,t) : \delta \Omega \times \Re_+ \rightarrow \Re$ and magnitude $r_{\text{sum}} \coloneqq r_{\text{sum}}(x,y,t) : \delta \Omega \times \Re_+ \rightarrow \Re_+$ form the polar coordinate with magnitude and direction equivalent to the average wave energy. This wave energy magnitude $\Psi_{\text{avg}_{\theta}} \coloneqq \Psi_{\text{avg}_{\theta}}(x,y,t) : \delta \Omega \times \Re_+ \rightarrow\Re$ is used to attenuate stokes drift velocity by the fraction of wave energy that is in the average wave direction $\Psi_{fr} \coloneqq \Psi_{fr}(x,y,t) : \delta \Omega \times \Re_+ \rightarrow \Re$, calculated by
	
	\begin{equation}
	\Psi_{fr} =\frac{ \Psi_{\text{avg}_{\theta}}}{\sum_{ i=1 }^{ k_x k_y } { \Psi_{\text{Total}_i} (k_x,k_y)} }.
	\end{equation}
	
	Stokes drift speed $\norm{\vec{U}_{{s}_z}}_2$ is in the direction of $\Psi_{\text{avg}_{\theta}}$, 	where $\vec{U}_{{s}_z} : \Omega \times \Re_+ \rightarrow \Re ^3$ is the stokes drift velocity vector $\vec{U}_{{s}_z}=[u_{s_z},v_{s_z},0]^T$, forming a stokes drift velocity:
	
	\begin{equation}\label{eq:stokesdriftprofile}
	\vec{U}_{{s}_z} = \omega k a_{p}^{2} e^{-2 k z} \Psi_{fr}.
	\end{equation}
	
	\subsection{Entrainment and buoyancy} \label{oilentrainment}
	Oil entrainment from the surface slick to the water column represents the movement of oil particles underwater by wave action and can be modelled as a random process with a probability for a particle to be entrained at a given time. The principle variable in the volume of oil entrained is the rate-scale scalar $\lambda_{\mathrm{o}\mathrm{w}} \coloneqq \lambda_{\mathrm{o}\mathrm{w}}(x,y,t) : \delta \Omega \times \Re_+ \rightarrow\Re$ \parencite{Tkalich2002}, which is defined by
	
	\begin{equation}\label{eq:entrainmentratescale}
	\lambda_{\mathrm{o}\mathrm{w}} = \frac{\pi k_e \gamma H_s}{8 \alpha T_{\text{peak}} L_{\mathrm{o}\mathrm{w}}},
	\end{equation}
	\noindent
	where $k_e \in [0.3,0.5] \subset \Re$ is an empirical constant, $H_s$ is the peak significant wave height, $T_{\text{peak}}$ is the wave period from the linear wave model and $L_{\mathrm{o}\mathrm{w}} \in \Re_+$ is a vertical length scale parameter that depends on the type of breaking wave. This is valued between 10m and 20m \parencite{Tkalich2002}. The vertical mixing term coefficient is $\alpha \in \Re \; : \; 1.15 \le \alpha \le 1.85$. The parameter~$\gamma \coloneqq \gamma(x,y,t) : \delta \Omega \times \Re_+ \in \Re$ is a dimensionless damping coefficient that takes the following values:
	\begin{center}
		$\gamma =\begin{cases} 10^5 \omega E_{\mathrm{w}}^{0.25}, \text{   for white-capping waves,} \\ 1.8 \times 10^{-7}\omega^{3}, \text{  for swell decay,} \end{cases}$
	\end{center}
	where $E_{\mathrm{w}} \coloneqq E_{\mathrm{w}}(x,y,t) : \delta \Omega \times \Re_+ \rightarrow \Re_+$ is calculated by
	\begin{equation}
	E_{\mathrm{w}} = \frac{g \rho_{\text{water}} H_s^{2}}{16},
	\end{equation}
	\noindent
	where $g$ is the gravitation acceleration constant of $9.81$m/s. The probability of entrainment~$P_\mathrm{s}$ for a Lagrangian particle for a discrete time-step $\Delta t \in \Re$ is as follows \parencite{Wang2010}:
	\begin{equation}\label{eq:probentrain}
	P_s = 1 - e^{(-\lambda_{\mathrm{o}\mathrm{w}} \Delta t)}.
	\end{equation}

	If the particle is inserted at this time-step, it enters the water column with intrusion depth:
	\begin{equation}\label{eq:entraindepth}
	D_i = (1.35 + 0.35(2 \phi - 1))H_s,
	\end{equation}
	\noindent
	where $\phi$ is a particle specific random variable with a uniform distribution in $[0,1]$ \parencite{Delvigne1988}. The maximum depth of intrusion, when $\phi = 1$, can be utilised as $z_{\text{crit}}$ to ensure a high resolution grid for entrained sub-surface oil particles.
	
	Oil particle buoyancy follows a typical scheme of instantaneous rising at a steady buoyancy velocity, determined by the oil droplet size, the water viscosity and the density difference \parencite{Proctor1994}. This buoyancy velocity is added to $w$, the vertical component of oil particle velocity. 
	
	\subsection{Thickness and mechanical spreading} \label{oilthickness}
	Following the advection, diffusion and entrainment of oil particles, additional particle movement is needed to represent the mechanical spreading of oil above its terminal spreading thickness.
	
	The volume of oil in each thickness map cell is calculated by summing the particles present in the cell, to form $V_{\text{oil}} \coloneqq V_{\text{oil}}(x,y,t) : \delta \Omega \times \Re_+ \rightarrow \Re_+$, in units of barrels for the empirical equation. This is then used to calculate area in square meters in Lehr's modified fay-type spreading formula \parencite{Lehr1984}, using the lower coefficient for a low wind case (as wind drift is accounted for elsewhere) and the average age of the oil in that cell $t_{\text{oil}} \coloneqq t_{\text{oil}}(x,y,t) : \delta \Omega \times \Re_+ \in \Re_+$ in minutes from the spill start. The empirical slick area $A_{\text{oil}} \coloneqq A_{\text{oil}}(x,y,t) : \delta \Omega \times \Re_+ \rightarrow \Re_+$ is found by computing
	
	\begin{multline}\label{eq:oilareacomplex}
	A_{\text{oil}} = 10^3 \biggl( 2.27 \frac{\rho_{\text{water}} - \rho_{\text{oil}}}{\rho_{\text{oil}}}^{\frac{2}{3}} V_{\text{oil}}^{\frac{2}{3}} t_{\text{oil}}^{-\frac{1}{2}} \\ + 0.03 \frac{\rho_{\text{water}} - \rho_{\text{oil}}}{\rho_{\text{oil}}}^{\frac{1}{3}} V_{\text{oil}}^{\frac{1}{3}} \norm{\vec{U}_{\mathrm{w}_{\text{knots}}}}^{\frac{4}{3}}_{2} t_{\text{oil}} \biggr),
	\end{multline}
	\noindent
	where $\vec{U}_{\mathrm{w}_{\text{knots}}}$ is the wind velocity converted to knots and where the oil age in hours $t_{\text{oil}}$ is capped to 48 hours, at which point mechanical spreading is minimal. Slick thickness in meters $\Gamma \coloneqq \Gamma(x,y,t) : \delta \Omega \times \Re_+ \rightarrow \Re_+$ in the grid cell of area $A_{\text{oil}}$ is then calculated as
	
	\begin{equation}
	\Gamma = \frac{V_{\text{oil}_{\text{m}^3}}}{A_{\text{oil}}},
	\end{equation}
	\noindent
	where $V_{\text{oil}_{\text{m}^3}}$ is the volume of oil in the cell converted to cubic meters. Depending on the oil type, if this thickness exceeds that of the equilibrium, or terminal oil thickness then mechanical spreading is applied using Lardners Lagrangian method in the local cell \parencite{Lardner2000}:
	
	\begin{subequations}\label{eq:mechanicalspreading}
		
		\begin{equation}
		\delta Q = 1.13 \left(\frac{\rho_{\text{water}} - \rho_{\text{oil}}}{\rho_{\text{water}}}\right)^{\frac{1}{3}} V_{\text{oil}_{\text{m}^3}}^{\frac{1}{3}} \frac{1}{4} t_{\text{oil}_\text{sec}}^{-\frac{3}{4}} \delta t,
		\end{equation}
		
		\begin{equation}
		\delta R = \delta Q + 0.0034 \norm{U_{\text{wind}_{10}}}^{\frac{4}{3}}_2 \frac{3}{4} t_{\text{oil}_\text{sec}}^{-\frac{1}{4}} \delta t,
		\end{equation}
		
		\begin{equation}
		x_{\text{new}} = x_0 + \delta Q \cos(\theta_{\text{wind}}) + \delta R \sin(\theta_{\text{wind}}),
		\end{equation}
		
		\begin{equation}
		y_{\text{new}} = y_0 + \delta Q\sin(\theta_{\text{wind}}) + \delta R \cos(\theta_{\text{wind}}).
		\end{equation}
		
	\end{subequations}
	
	For this empirical formula, $t_{\text{oil}}$ is in seconds and $\theta_{\text{wind}} \coloneqq \theta_{\text{wind}}(x,y,t) : \delta \Omega \times \Re_+ \rightarrow \Re$ is the wind angle, or bearing from north of $\vec{U}_{\mathrm{w}}$. The distances $\delta Q \in \Re$ and $\delta R \in \Re$ represent the mechanical spreading and the augmented mechanical spreading from wind effects respectively. Equations within Sections \ref{oiladvectionanddiffusion}, \ref{oilentrainment} and \ref{oilthickness} have described the movement of oil particles in the surface and subsurface ocean, but have not accounted for any changing in oil properties through weathering or particle deposition on obstacles and shorelines.
	
	\subsection{Oil deposition}
	The model currently assumes zero particle movement once it enters a non-water cell. If the beach cell is considered saturated, the particle cannot enter \parencite{Chao2003} and remains afloat. This offers simple shore deposition, though particles cannot re-float once deposited.
	
	\subsection{Number of oil particles}\label{sec:numberofoilparticles}
	The presence of random processes modelling oil turbulent diffusion and entrainment cause the spreading of oil particles to become a stochastic process in the simulation. Therefore the number of particles required in the simulation is not determined by the need for accurate reconstruction of a spill shape, but by the need to adequately sample the combined probability distribution function to resolve the process. The stochastic element is a combination of a 2D random walk, a 1D random walk and a dichotomous binomial distribution with a uniform distribution. These are horizontal turbulent diffusion, vertical turbulent diffusion and binary entrainment at a uniformly random depth.
	
	First consider the horizontal turbulent diffusion random walk: Although the distance from origin is not accurately represented by a Normal distribution as samples cannot take values less than zero, the distribution of particles along an individual axis can be assumed Normal. The horizontal diffusion Normal distribution has the parameters
	
	\begin{subequations}
		\begin{equation}
		\sigma_\text{horz} = (\sqrt{2}-1) \sqrt{12 D_h \delta t}
		\end{equation}
		and
		\begin{equation}
		\mu_\text{horz} = 0,
		\end{equation}
	\end{subequations}
	forming the distribution $N(\mu_\text{horz},\sigma_\text{horz}^2)$. Define the confidence interval $\alpha_\text{horz} \in \Re$ and expected random walk movement $E_\text{horz} \in \Re$ by
	
	\begin{subequations}
		
		\begin{equation}
		\alpha_\text{horz} = 0.05,
		\end{equation}
		and
		\begin{equation}
		E_\text{horz} = \frac{1}{2}\sqrt{12 D_h \delta t},
		\end{equation}
		
	\end{subequations}
	then the number of samples needed to approximate the random walk process with a $95$\% confidence level is \parencite{NIST/SEMATECH2012}:
	
	\begin{equation}\label{eq:samplesneeded}
	n_\text{horz} \ge \left(\frac{1.96}{\alpha_\text{horz}E_\text{horz}}\right)^2 \sigma_\text{horz}^2. 
	\end{equation}
	
	Under typical simulation conditions in Beaufort scale 5 sea states, $n_\text{horz}$ in \eqref{eq:samplesneeded} has a value of approximately $1000$, which exceeds the samples needed to approximate the vertical turbulent diffusion, uniform entrainment depth and the number of samples required to apply the central limit theorem to the dichotomous binomial distribution of entrainment. Given the complex interaction between stochastic processes that would greatly increase the variance of the combined probability function, the negligible effect on computational time of increased numbers of oil particles and the implicit desire to improve the simulation accuracy and confidence limit, it is recommended that a minimum of $3000$ particles be used. This also exceeds the sum of sample sizes needed for each random process in typical conditions.

	\subsection{The probability of oil presence}\label{sec:oilprobability}
	In a simulation realization identified by $S_n \in \mathbb{N}$, the presence of oil at a time-step $t_k \in \Re_+$ in the cell at position $(x_i,y_j,z_w) \in \Omega$ is described by the binary random variable $O_p(x_i,y_j,z_w,t_k,S_n)$, which takes the value $0$ when the oil volume in the cell at $(x_i,y_j,z_w)$ is less than an arbitrary threshold value $\zeta_p \in \Re_+$ (no oil present) and the value $1$ when the oil volume in the cell is greater than $\zeta_p$ (oil is present), at time-step $t_k$. The binary random variable is described by 
	
	\begin{equation}
	O_p(x_i,y_j,z_w,t_k,S_n) =
	\begin{cases}
	\begin{aligned}
	0 &\text{ when } \tilde{V}_{\text{oil}}(x_i,y_j,z_w,t_k,S_n) \leq \zeta_p, \\
	1 &\text{ when } \tilde{V}_{\text{oil}}(x_i,y_j,z_w,t_k,S_n) > \zeta_p,
	\end{aligned}
	\end{cases}
	\end{equation}
	
	\noindent
	where the function $\tilde{V}_{\text{oil}}(x_i,y_j,z_w,t_k,S_n) : \Omega \times \Re_+ \times \mathbb{N}_+ \rightarrow \Re_+$ returns the volume of oil present in the discrete cell $(x_i,y_j,z_w)$ at time $t_k$ for realization $S_n$. Consequently, the evolution of oil presence across the spatial domain is described by the stochastic process $\{O_{p}(\Omega,t_k,S_n) \}_{t_k \in \Re_+}$, the set of binary random variables describing oil presence in the spatial domain $\Omega$ for each time-step $t_k$, for realization $S_n$. The presence of oil in a set of space and time $A \subseteq \Omega \times \Re_+$ that may span multiple time steps, on a discrete mesh, is characterised by the binary random variable
	
	\begin{equation}
	\tilde{O}_p (A,S_n) =
	\begin{cases}
	\begin{aligned}
	0 &\text{ when } \smashoperator[lr]{\sum_{x_i,y_j,z_w,t_k \in A}^{}} O_p(x_i,y_j,z_w,t_k,S_n) = 0, \\
	1 &\text{ when } \smashoperator[lr]{\sum_{x_i,y_j,z_w,t_k \in A}^{}} O_p(x_i,y_j,z_w,t_k,S_n) \neq 0.
	\end{aligned}
	\end{cases}
	\end{equation}
	
	\noindent
	Hence, $\tilde{O}_p (A,S_n)$ only takes value 0 if the volume oil in every cell is less than or equal to $\zeta_p$ for the entire spatio-temporal set $A$, or takes value 1 if the oil volume in any cell exceeds $\zeta_p$ at any time, in the realization $S_n$.
	
	To inform sensor placement it is useful to describe the probability of oil presence in $A$, by utilising multiple realizations each of which is assumed to be an independent stochastic process. Multiple realizations are needed to examine model sensitivity to uncertain parameters, such as drift coefficients. The probability of oil volume exceeding $\zeta_p$ using $S_t \in \mathbb{N}$ realizations, $\text{P}(\hat{O}_{p}(A) = 1,S_t)$ is defined by
	
	\begin{equation}\label{eq:probabilityoilpresencemultiplesimulations}
	\text{P}(\hat{O}_{p}(A) = 1,S_t) = \frac{\sum_{S_n = 1}^{S_n = S_t}{\tilde{O}_p(A,S_n)}}{S_t}.
	\end{equation}
	
	To determine the number of realizations needed to adequately sample the random processes, the probability of oil presence sample variance after $S_t$ realizations is calculated \parencite{Montgomery1994} by
	
	\begin{subequations}
		\begin{multline}\label{eq:variance}
		\text{Var}(\text{P}(\hat{O}_{p}(A) = 1)) =\\ \frac{1}{S_t - 1}\sum_{S_n = 1}^{S_n = S_t}\left(\text{P}(\hat{O}_{p}(A) = 1,S_n) - \bar{\text{P}}(\hat{O}_{p}(A) = 1),S_n)\right)^{2},
		\end{multline}
		
		\noindent
		with a maximum value across $\Omega$ of
		
		\begin{equation}\label{eq:variancemax}
		\text{Var}_{\text{max}}(\text{P}(\hat{O}_{p}(A) = 1)) = \max_{A \in \Omega} \left(\text{Var}(\text{P}(\hat{O}_{p}(A) = 1)\right),
		\end{equation}
		
	\end{subequations}
	\noindent
	where $\bar{\text{P}}(\hat{O}_{p}(A) = 1,S_t) = \frac{1}{S_t} \sum_{S_n = 1}^{S_n = S_t} \text{P}(\hat{O}_{p}(A) = 1,S_n)$ is the mean probability of oil presence for $S_t$ realizations. For the parameters of Table \ref{tab:americagrandeparameters} and an oil threshold value of $\zeta_p = 0$, the maximum value of the variance \eqref{eq:variancemax} with realization number decreases rapidly, then settles after $S_n \approxeq 200$ as in Figure \ref{fig:oilprobvar}. The variance distribution of \eqref{eq:variance} displayed peaks at the trail and leading edges of the spill, as expected due to the changing in presence of oil across realizations compared to the overlap of spills at the spill centre. The variance in oil presence probability is used instead of the variance in oil presence, as a confidence interval in oil probability describes a range of chance in oil presence and is more useful than a confidence interval in number of realizations with oil present.
	
	\subsection{The probability of oil particle drift location at a specific time}
	Another useful event to model is the surface location of a selected oil volume at a given time-step. Define the vector valued random variable $O_v(p_i,t_k,S_n) = [x_p,y_p] \in \delta \Omega$ to take the vector position of the oil particle index $p_i \in \mathbb{N}$ at the time-step $t_k$, with $x_p \in \Re$ and $y_p \in \Re$ the horizontal and vertical locations respectively, for a given realization $S_n$. The probability of an oil particle $p_i$ to be within the discrete cell $(x_i,y_j)$ at $t_k$, $\text{P}(O_v(p_i,t_k,S_n) \in (x_i,y_j))$, is defined by
	
	\begin{equation}\label{eq:discreteoilprobabilityparticles}
	\text{P}(O_v(p_i,t_k,S_n) \in (x_i,y_j)) = \frac{\smashoperator[r]{\sum\limits_{p_i \in p_p(x_i,y_j,t_k,S_n)}^{}} V_{\text{particle}}(p_i,S_n)}{\smashoperator[r]{\sum\limits_{p_i \in p_t(t_k,S_n)}^{}} V_{\text{particle}}(p_i,S_n)},
	\end{equation}
	
	\noindent
	where $p_p(x_i,y_j,t_k,S_n) : \delta \Omega \times \Re_+ \rightarrow \mathbb{N}^{m_p}$ is a vector of particle indices present in the discrete spatio-temporal location and $p_t(t_k,S_n) \rightarrow \mathbb{N}^{m_t}$ is a vector of all particle indices at time $t_k$, with $m_p$ and $m_t$ being the number of oil particles present and the total number of oil particles respectively. The oil volume function $V_{\text{particle}}(p_i,S_n) : \mathbb{N}^{m_t} \rightarrow \Re_+$ maps oil particle indices $p_i$ to the oil volume they represent in the model. Evaluation of \eqref{eq:discreteoilprobabilityparticles} for every cell in $\delta \Omega$ forms the probability mass function displayed in Figures \ref{fig:oilprob19} and \ref{fig:oilprob23}. The probability of oil presence in cell $(x_i,y_j)$ is obtained by averaging over the realizations of the stochastic process. The resulting probability is given by

	\begin{multline}\label{eq:probabilitymultiplesimulations}
	\text{P}(\hat{O}_v(p_i,t_k) \in (x_i,y_j)) = \\ \frac{1}{S_t} \sum_{S_n=1}^{S_n=S_t} \text{P}(O_v(p_i,t_k,S_n) \in (x_i,y_j)),
	\end{multline}
	
	\noindent
	where $\text{P}(O_v(p_i,t_k,S_n) \in (x_i,y_j))$ is the evaluation of \eqref{eq:discreteoilprobabilityparticles} for that realization index. This probability, $\text{P}(\hat{O}_v(p_i,t_k) \in (x_i,y_j))$, provides a further measure for route planning by indicating likely areas of high oil volume, while the probability of oil presence $\text{P}(O_p(x_i,y_j,z_w,t_k)=1)$ defines likely areas of any oil exceeding a threshold $\zeta_p$. 
	
	\subsection{The mean location of the spill centre}
	A further vector valued random variable $O_{m}(t_k,S_n) = [x_m,y_m] \in \delta \Omega$ takes the value of the position of highest oil volume for realization $S_n$, where $x_m \in \Re$ and $y_m \in \Re$ are the horizontal and vertical locations of the highest volume position respectively. The value taken by $O_{m}(t_k,S_n)$ is one definition of the spill centre. Define the mean spill centre position across realizations by
	
	\begin{equation}\label{eq:oilmeancentre}
	\bar{O}_{m}(t_k) = \frac{1}{S_t} \sum_{S_n = 1}^{S_n = S_t} O_{m}(t_k,S_n).
	\end{equation}
	\noindent
	The mean spill centre position for 500 realizations is displayed in Figure \ref{fig:oilprobmonte}.
	
	\section{Model simulation and results}\label{results}
	The model is intended to guide sensing assets in the aftermath of maritime incidents and hence requires validation, with comparison against real-world data preferable \parencite{Spaulding2017}. The Grande America oil spill of March 2019 provides a recent and observed incident to validate against. However due to the vessel's abandonment on the 11th March 2019 due to an onboard fire and the subsequent sinking in water depth of 4600m between 1500 and 1800 hours on the 12th March 2019, it is unclear exactly when the vessel sank, the oil leak occurred, or how much leaked. This information forms the initial conditions for the spill and can heavily affect simulation results.
	
	For the model simulation it is assumed the fuel tanks became compromised as the hull split and sank and the worst case scenario is modelled: all 2200 tonnes of Heavy Fuel Oil carried by the Grande America is spilt in a short time-frame, from 1400 to 1600 hours on the 12th March 2019, at coordinates -5.7844$^\circ$ East, 46.0689$^\circ$ North. The model utilises Global Forecast System (GFS) wind velocities and Tide-Tech ocean velocities, with a North-West to South-East wave swell with significant wave height of 3m from National Centers for Environmental Prediction (NCEP) data. User specified parameters are presented in Table \ref{tab:americagrandeparameters}. The Grande America oil spill was observed by the Copernicus Sentinel 1 and 2 satellites on two occasions, on the 19th March 2019 the 5 day old slick is observed at approximately 45.439458$^\circ$ North, -4.283424 $^\circ$ East and on the 23rd March 2019 the 11 day old slick is observed at 45.0826$^\circ$ North, -4.4559$^\circ$ East.
	
\begin{figure*}
		\centering
		\begin{minipage}{.48\textwidth}
			\centering
			\includegraphics[trim={2cm 0 0 0},clip,width=1.1\linewidth]{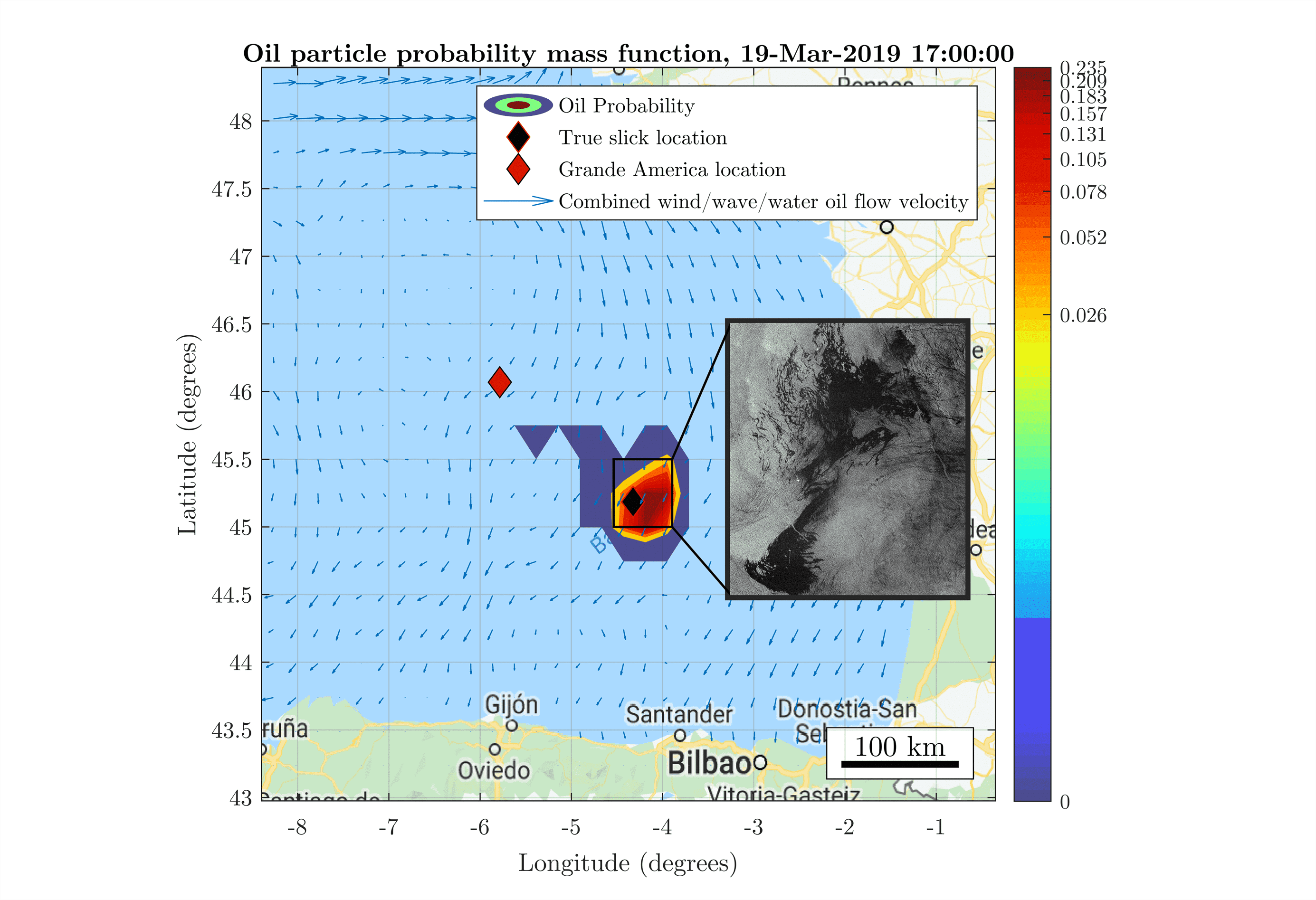}
			\captionof{figure}{The probability mass function of oil particle drift \eqref{eq:discreteoilprobabilityparticles} from the SCEM simulation for 17:00 on 19th March 2019, 5 days after the spill released, using a log scale and with the real position marked. Note the similarity in location to the real slick location on the 19th March. Map data \textcopyright 2019 Google, Inst. Geogr. Nacional. Contains modified Copernicus Sentinel data (2019), processed by ESA,	\href{https://creativecommons.org/licenses/by-sa/3.0/igo/}{CC BY-SA 3.0 IGO}.}
			\label{fig:oilprob19}
		\end{minipage}%
		\hfill
		\begin{minipage}{.48\textwidth}
			\centering
			\includegraphics[trim={2cm 0 0 0},clip,width=1.1\linewidth]{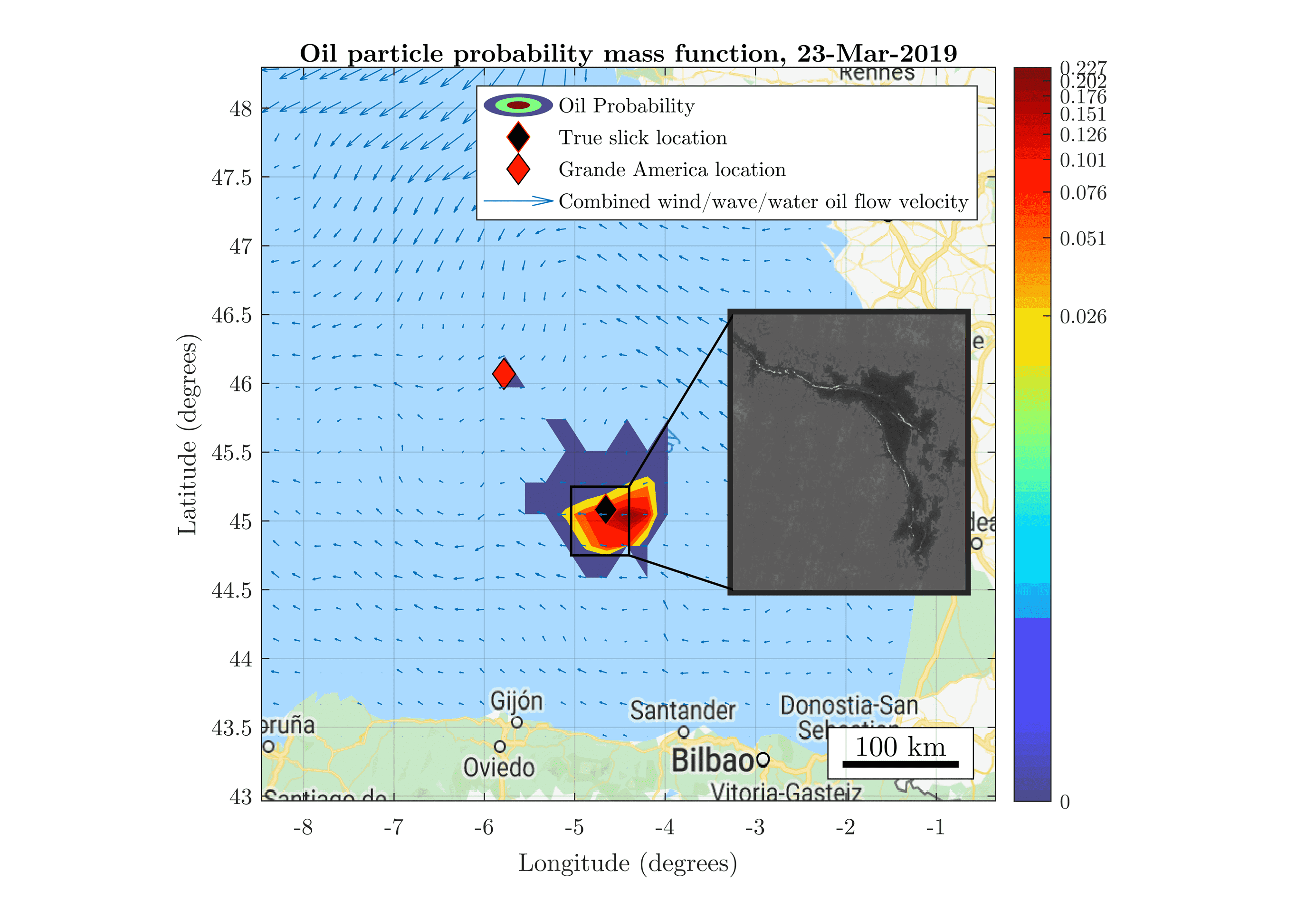}
			\captionof{figure}{The probability mass function of oil particle drift \eqref{eq:discreteoilprobabilityparticles} from the SCEM simulation for the 23rd March 2019, 11 days after the spill released, using a log scale and with the real position marked. Note the similarity in location to the real slick location on the 23rd March. Map data \textcopyright 2019 Google, Inst. Geogr. Nacional. Contains modified Copernicus Sentinel data (2019), processed by ESA,	\href{https://creativecommons.org/licenses/by-sa/3.0/igo/}{CC BY-SA 3.0 IGO}.}
			\label{fig:oilprob23}
		\end{minipage}
	\end{figure*}

	Due to the uncertainty surrounding the initial spill conditions and volume, emphasis is placed on the model accurately predicting the drift of a spill, with little importance placed on predicting the slick thickness or volume. The probability of oil drift location in a spatio-temporal domain is given by equation \eqref{eq:discreteoilprobabilityparticles}, this is evaluated for each grid-cell area at the indicated time to produce Figures \ref{fig:oilprob19} and \ref{fig:oilprob23}. Figures \ref{fig:oilprob19} and \ref{fig:oilprob23} show accurate prediction of the slick locations, with high probability at 45.2000$^\circ$ North, -4.1850$^\circ$ East on the 19th March 2019, with the true location being 45.1857$^\circ$ North, -4.323424 $^\circ$ East, $\approx$14km to the north west. For the 23rd March 2019, with no correction or reinitialisation from the true spill position on the 19th March, the model predicts a slick location at 45.0300$^\circ$ North, -4.2100 $^\circ$ East, compared to the true position at 45.0826$^\circ$ North, -4.6559$^\circ$ East, 20km to the west of the predicted position. Errors of 15km and 20km for five and eleven day predictions, respectively, not unreasonable given the scale of the spill, the large size of the domain, the lack of model correction or calibration and the model's intended purpose for predictions over much shorter time-scales (hours to a day). The 288 hour prediction took 568 seconds to compute in \textsc{Matlab}, on a Windows 10, i7-6700k CPU desktop computer, this includes computation time for the wind, wave and ocean hydrodynamic models across 2688 surface nodes, extrapolated to 534912 sub-surface nodes. All parameters were within ranges acceptable to literature and use their values within Table \ref{tab:americagrandeparameters}.
	
\begin{figure*}
		\centering
		\begin{minipage}{.48\textwidth}
			\centering
			\includegraphics[trim={2cm 0 0 0},clip,width=1.1\linewidth]{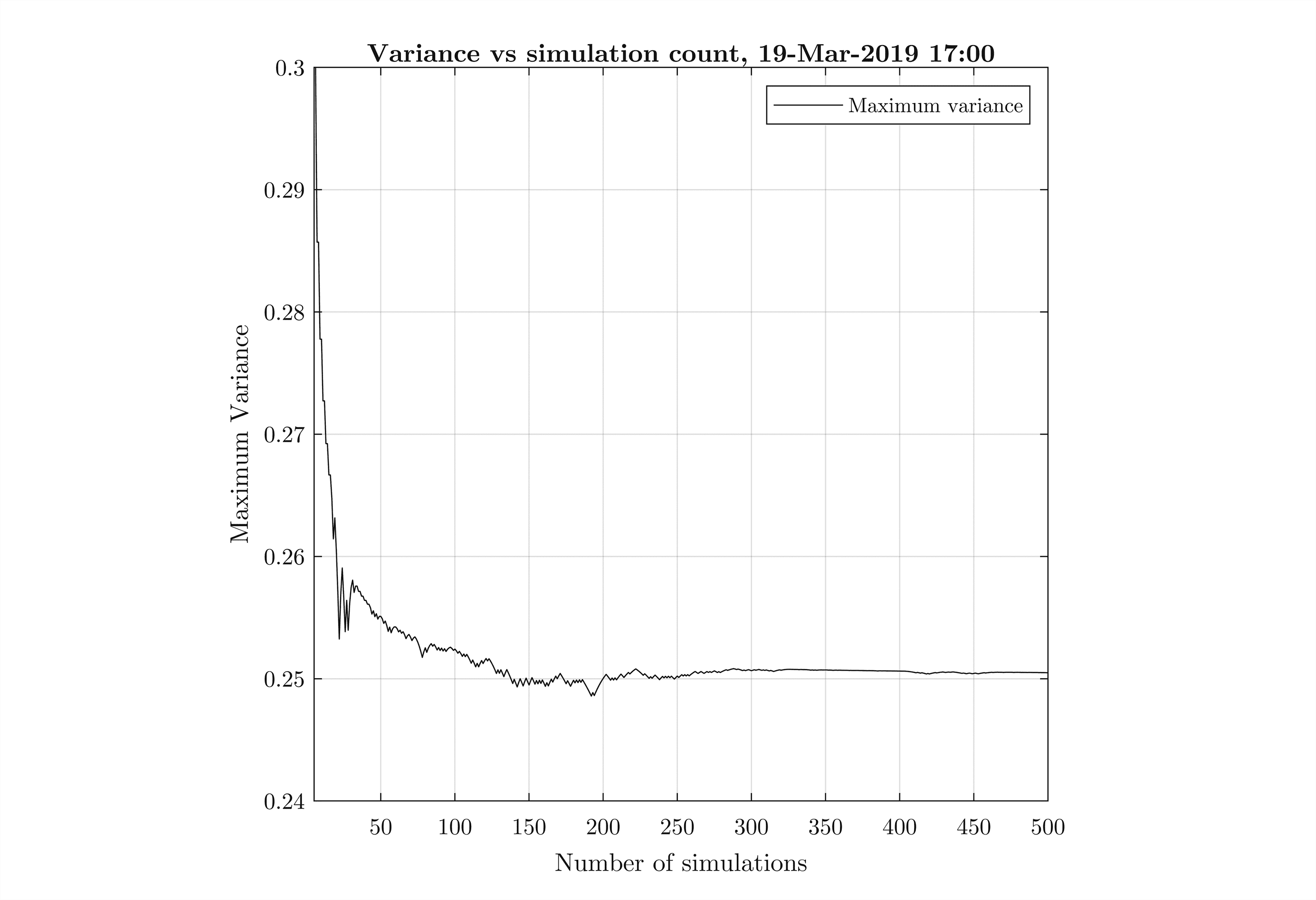}
			\captionof{figure}{The plot showing the decay of the maximum variance of oil presence \eqref{eq:variancemax} across 500 SCEM simulations for 17:00 on the 19th March 2019, 5 days after the spill released. Note the rapid decay and convergence, settling around 200 simulations. \\ \\ \\ \\ \\}
			\label{fig:oilprobvar}
		\end{minipage}
		\hfill
		\begin{minipage}{.48\textwidth}
			\centering
			\includegraphics[trim={2cm 0 0 0},clip,width=1.1\linewidth]{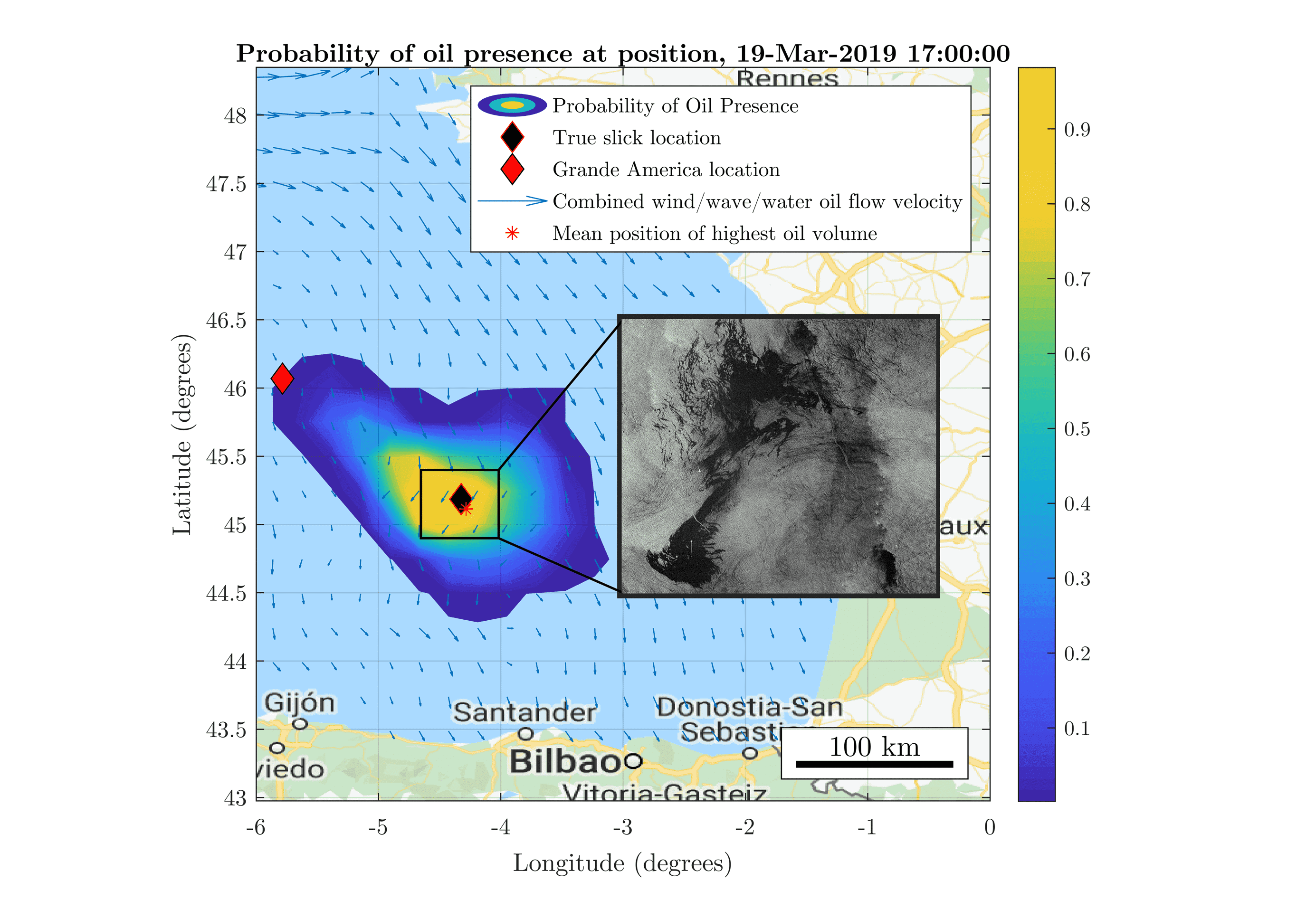}
			\captionof{figure}{The combined probability map of oil presence \eqref{eq:probabilityoilpresencemultiplesimulations} for 17:00 on the 19th March 2019, 5 days after the spill released, with the real position marked. Each SCEM simulation was a sampling of spill parameters in Table \ref{tab:americagrandeparameters}. Note the similarity in location of the highest probability and mean spill position to the real slick location on the 19th March. Map data \textcopyright 2019 Google, Inst. Geogr. Nacional. Contains modified Copernicus Sentinel data (2019), processed by ESA, \href{https://creativecommons.org/licenses/by-sa/3.0/igo/}{CC BY-SA 3.0 IGO}.}
			\label{fig:oilprobmonte}
		\end{minipage}%
	\end{figure*}

	To investigate the sensitivity to the spill parameters and the diffusion, wind and ocean current coefficients of equation \eqref{eq:oilmovement}, 500 simulation realizations using simultaneous sampling of the random variable coefficients of Table \ref{tab:americagrandeparameters} were utilised to get a probability of oil presence map \eqref{eq:probabilityoilpresencemultiplesimulations} across the set of random variables. Figure \ref{fig:oilprobmonte} shows that the model is accurate for the Grande America spill within the typical bounds for drift parameters and that the default coefficient values slightly overestimate oil movement up to the 19th March 2019 when comparing the results of Figures \ref{fig:oilprob19} and \ref{fig:oilprobmonte}.
	
	\begin{sidewaystable*}%[htbp]
		\centering
		\begin{tabular}{|r|l|l|l|l|}
			\hline
			\multicolumn{5}{|c|}{America Grande simulation parameters} \\
			\hline
			Symbol & Description & Value & Nominal distribution within bounds & Units \\
			\hline
			\hline
			\multicolumn{1}{|l|}{$\alpha_{\mathrm{w}_0}$} & Oil wind advection coefficient & 0.02 &$\alpha_{\mathrm{w}_0} \sim \text{U}(0.005,0.03)$ & - \\
			\hline
			\multicolumn{1}{|l|}{$\alpha_{\mathrm{c}_0}$} & Oil current advection coefficient & 1.0 & $\alpha_{\mathrm{c}_0} \sim \text{U}(0.9,1.1)$ & - \\
			\hline
			\multicolumn{1}{|l|}{$c_{\text{smag}}$} & Diffusion empirical parameter & 0.1 & $c_{\text{smag}} \sim \text{U}(0.01,0.3)$ & - \\
			\hline
			\multicolumn{1}{|l|}{$t_0$} & Spill leak start date & March 12th 2019 - 03:30:00 & $t_0 \sim \text{U}(\text{11/03/19 - 22:00:00, 12/03/19 - 17:00:00})$ & - \\
			\hline
			\multicolumn{1}{|l|}{$t_f$} & Spill leak end date & March 12th 2019 - 16:30:00 & $t_f \sim \text{U}(\text{12/03/19 - 12:00:00, 12/03/19 - 18:00:00})$ & - \\
			\hline
			\multicolumn{1}{|l|}{$V_{T}$} & Total spill volume & 2200 & $V_{T} \sim \text{U}(10, 2200)$ & Tonnes \\
			\hline
			\multicolumn{1}{|l|}{$n_x,n_y,n_z$} & Domain nodes & $64 \times 42 \times 200$ & - & - \\
			\hline
			\multicolumn{1}{|l|}{$\Omega$} & Domain size & $664.3 \times 443.0$ & - & km \\
			\hline
			\multicolumn{1}{|l|}- & Oil Properties & Heavy Fuel Oil (No. 6 Fuel oil) & - & - \\
			\hline
		\end{tabular}%
		\caption{\label{tab:americagrandeparameters} Tabulation of parameters used in the Grande America oil spill simulation.}
	\end{sidewaystable*}%
	
	As demonstrated by the Grande America spill, data is scarce on spill components, environmental data, contaminant position and thickness following a real incident, with even international scale incidents only becoming well observed and documented several days after the incident.
	
	Given that the model may be required to make predictions for spills in remote areas, where existing hydrodynamic models may be inaccurate or non existent, the model is now compared to the industry standard model GNOME as a benchmark to see if it provides similar predictions when provided with the same inputs. As a further comparison and as a pointer to future work, a test is performed to see if the model can offer similar results when given erroneous inputs (hydrodynamic data without tide flow in a river delta for example), by using real-time sensory feedback from sensors taking data from a twin GNOME simulation with accurate input data.
	
	This section also presents comparison of several spill simulations using this oil model and GNOME, utilising three sets of data, Global Forecast System (GFS) wind velocities, HYCOM ocean velocities (that contain no tidal flow data, only circulation flow) and Tide-Tech ocean velocities. As large data set acquisition is unlikely to be available in deployment due to data transmission constraints, both the HYCOM and Tide-Tech ocean data results are assumed to be available as surface velocity only. Both sets include Ekman currents and therefore Ekman currents are omitted from the surface dynamics, but a spiral is calculated sub-surface to model oil slick shear separation between surface and sub-surface particles. Both models utilise the same number of particles, representing the same volume of oil each and released at the same leak rate from the same location. Therefore, particle positions can be utilised for comparative purposes.
	
	\begin{figure*}
		\centering
		\begin{minipage}{.48\textwidth}
			\centering
			\includegraphics[trim={2cm 0 0 0},clip,width=1.1\linewidth]{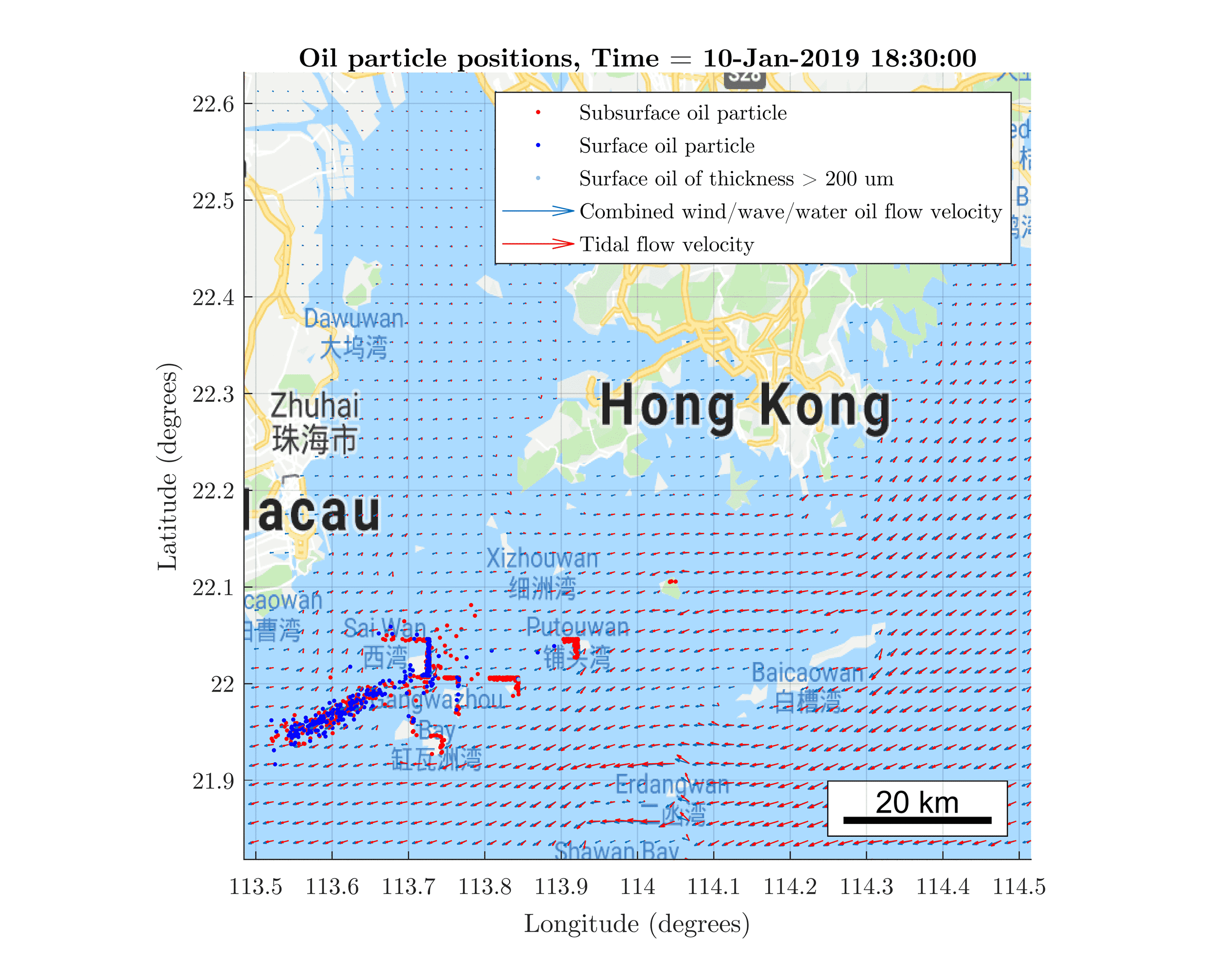}
			\captionof{figure}{The model and GNOME simulation results for a 3-day simulation of a 100 barrel spill released 1 mile south of Lamma Island, Hong Kong, at 0330 hours on the 8th January 2019. Both models has been forced by GFS wind velocities and HYCOM ocean velocities. Note the presence of oil on all the same islands and positions of the leading edge of the spill. Map data \textcopyright 2019 Google.}
			\label{fig:experiment12}
		\end{minipage}%
		\hfill
		\begin{minipage}{.48\textwidth}
			\centering
			\includegraphics[trim={2cm 0 0 0},clip,width=1.1\linewidth]{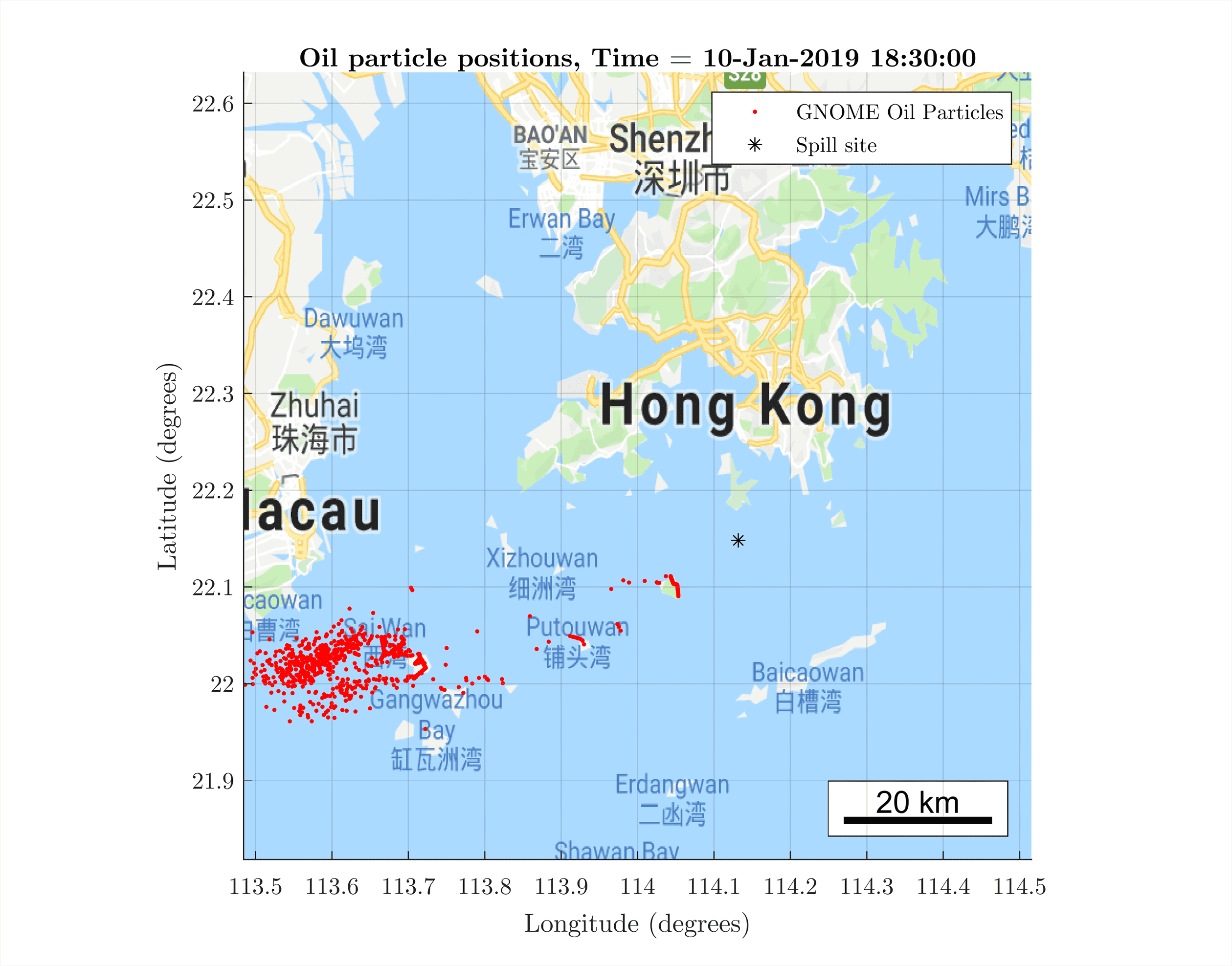}
			\captionof{figure}{GNOME results for a 3-day simulation of a 100 barrel spill released 1 mile south of Lamma Island, Hong Kong, at 0330 hours on the 8th January 2019. GNOME has been forced by GFS wind velocities and Tide-Tech ocean velocities that include tidal flow. Map data \textcopyright 2019 Google.}
			\label{fig:experiment21}
		\end{minipage}
	\end{figure*}
	
	Experiment 1 is a 3-day simulation of a 100 barrel spill released 1 mile south of Lamma Island, Hong Kong, at 0330 hours on the 8th January 2019 carried out as a contingency for the Aulac Fortune oil tanker explosion. The oil models are both forced by HYCOM ocean current data and GFS wind data and their similarity displayed in Figure \ref{fig:experiment12} validates the resolving of external data in the environment model (Section \ref{environmentmodel}) and oil model (Section \ref{oilmodel}).
	
	Experiment 2 demonstrates the sensitivity of oil models to the environment model they are driven by. The GNOME model has been driven by the GFS wind data and by Tide-Tech ocean current data sets that include a harmonic tide component reflecting the important effect of the Zhujiang river estuary. The raw Tide-Tech data has been modified to a higher resolution and resolved around islands using the Navier-Stokes simulation of Section \ref{environmentmodel}, with original values preserved where possible. Note the large discrepancies between the GNOME results of Figure \ref{fig:experiment21} and Figure \ref{fig:experiment12}, with the majority of flow passing to the north of the island cluster at 22N 113.7E when tidal flow is present.

	\begin{figure*}
		\centering
		\makebox[\textwidth][c]{\includegraphics[width=\textwidth]{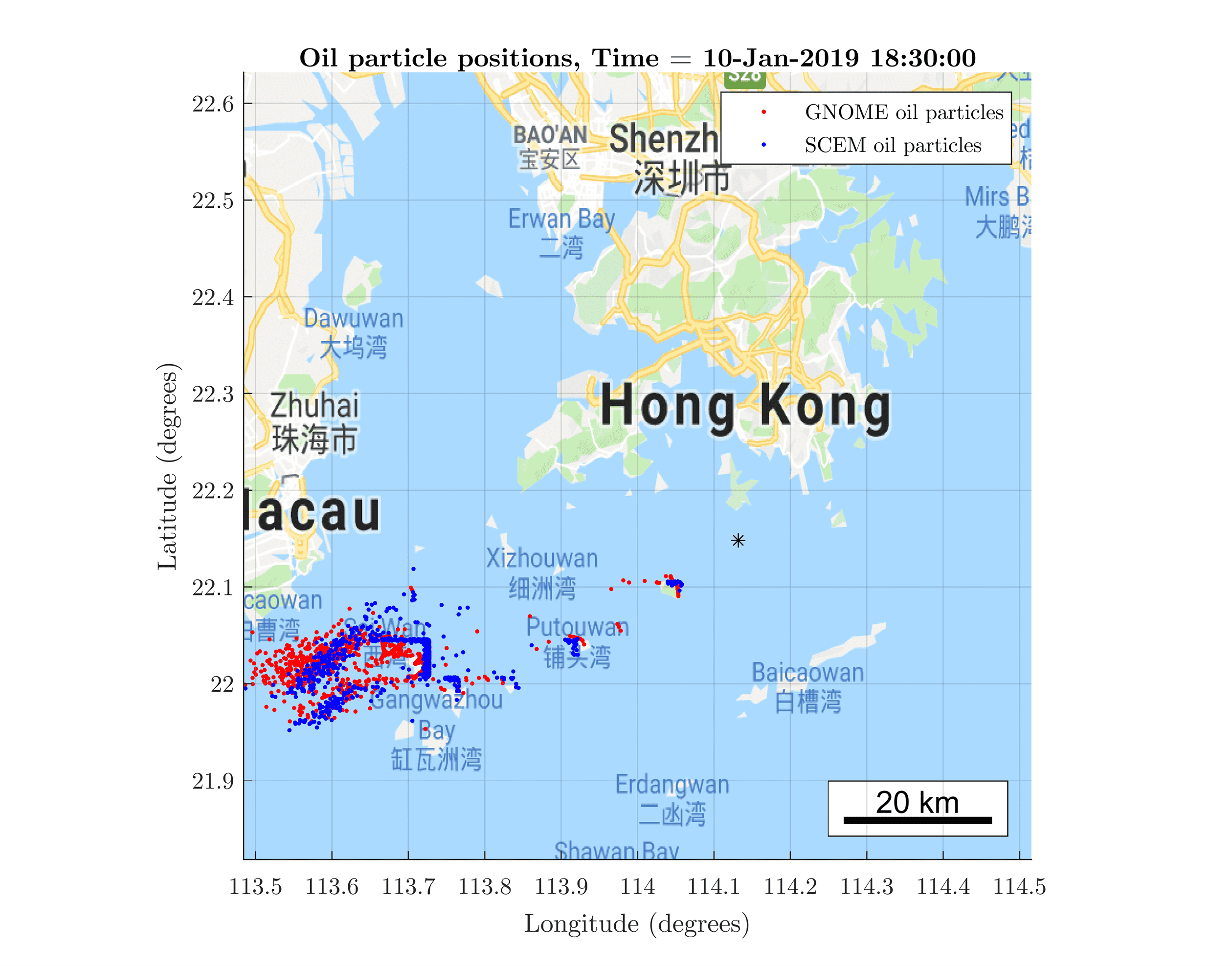}}
		\caption{\label{fig:experiment32} The model and GNOME simulation results for a 3-day simulation of a 100 barrel spill released 1 mile south of Lamma Island, Hong Kong, at 0330 hours on the 8th January 2019. The model has been forced by GFS wind velocities and HYCOM ocean velocities, while GNOME has been forced by GFS wind velocities but Tide-Tech ocean velocities that include tidal flow. Note the presence of oil on all the same islands, concentration and positions of the leading edge of the spill. Map data \textcopyright 2019 Google.}
	\end{figure*}
	
	Experiment 3 demonstrates the effectiveness of sensors when informing the model described here-in, to deliver accurate results even when prior information is inaccurate. The model is forced by HYCOM ocean current data (that lack strong tidal components) and GFS wind data, but well-guided mobile sensors capable of measuring wave, wind and current properties and oil thickness are capable of minimising the error in oil estimation that results from external data differences. The methodology utilises SCEM and is detailed in future work. The four sensors measure flow and oil values at point locations from the GNOME Experiment 2 run, at regular intervals of every 15 minutes, and are constrained by their maximum speed of 50mph. It is assumed the sensors are capable of measuring the flow properties perfectly and their measurements inform a time-varying Kalman Filter utilising a modal decomposition model of SCEM, to estimate SCEM states. Figure \ref{fig:experiment32} displays the similarity between the spill particles of the model and those of the GNOME simulation, with both the spills leading edge and dominant concentrations matching, including having a majority flow around the north of the island cluster at 22N 113.7E, despite tidal flow being absent from input data in this model.
	
	\section{Conclusion}
	Motivated by the need to monitor environment properties or pollutants in the aftermath of maritime incidents, a model-based adaptive monitoring strategy is developed for the the emerging application of mobile sensors. This paper presents the Sheffield Combined Environment Model (SCEM), the environment and oil model component of the monitoring methodology used in Figure \ref{fig:experiment32}, for the purpose of providing online control guidance to assets with minimal supporting data. The new model is described, giving equations and algorithm in both flow chart (Figure \ref{fig:modeldiagram}) and pseudocode (Algorithm \ref{fig:algorithm}). The model is then demonstrated to accurately predict a real-world oil spill in the Bay of Biscay 2019 and give similar results to an industry standard oil model GNOME when given the same input data for a spill near Hong Kong 2019.
	
	Furthermore, a sensing strategy developed using SCEM is shown to be capable of delivering an accurate estimation of oil positions when given inaccurate external forcing data. This provokes further research into optimal sensor placement and sensor feedback methods through application of optimisation, estimation and data fusion fields of work.

	\printbibliography
	
\end{document}